\documentclass[twocolumn,prb]{revtex4}

\usepackage{graphicx}
\usepackage{dcolumn}
\usepackage{bm}

\begin{document}

\title{
Critical charge dynamics of superconducting La$_{2-x}$Sr$_x$CuO$_4$ thin films 
probed by complex microwave spectroscopy:\\
Anomalous changes of the universality class by hole doping}

\author{T. Ohashi}
 \affiliation{
	Department of Basic Science, The University of Tokyo,\\
	3-8-1, Komaba, Meguro-ku, Tokyo 153-8902, Japan
 }

\author{H. Kitano}
 \affiliation{
	Department of Physics and Mathematics, Aoyama Gakuin University,\\
	5-10-1 Fuchinobe, Sagamihara, Kanagawa 229-8558, Japan
 }

\author{I. Tsukada}
 \affiliation{
	Central Research Institute of Electric Power Industry,\\
	2-11-1, Iwadokita, Komae, Tokyo 201-8511, Japan
 }

\author{A. Maeda}
 \affiliation{
	Department of Basic Science, The University of Tokyo, \\
	3-8-1, Komaba, Meguro-ku, Tokyo 153-8902, Japan
 }

\date{\today}

\begin{abstract}
We study the critical charge dynamics of the superconducting to the normal-state transition 
for La$_{2-x}$Sr$_x$CuO$_4$ (LSCO) thin films with a wide range of the Sr concentration, 
by measuring the frequency-dependent excess parts of the complex microwave conductivity, 
which is induced by the superconducting fluctuations. 
We present a dynamic scaling analysis of the complex fluctuation conductivity, 
which includes the information on the universality class 
and the dimensionality of the critical charge dynamics as a function of 
the Sr concentration, the film thickness and the magnetic field. 
In our previous study (H. Kitano {\it et al.}, Phys. Rev. B {\bf 73}, 092504 (2006).),
the 2D-$XY$ critical dynamics for underdoped LSCO 
and the 3D-$XY$ critical dynamics for optimally doped LSCO were reported.
In this study, we observed a novel two-dimensional unknown critical charge dynamics 
for overdoped thin films from $x$=0.17 to 0.20, 
which is clearly distinguished from the 2D-$XY$ critical dynamics. 
Through the systematic measurements by changing the film thickness 
or by applying small magnetic field, 
it was confirmed that this unusual behavior, which is referred as 2D-``U" below, 
was not induced by the finite size effect but was intrinsic to the overdoped LSCO, 
Thus, it was found that the critical behavior in the phase diagram of LSCO 
is classified into the following three types; 
(i) 2D-$XY$ for underdoped region, 
(ii) 3D-$XY$ for optimally doped region, 
and (iii) 2D-``U" for overdoped region. 
In other words, the dimensionality in the critical charge dynamics 
is changed twice with hole doping. 
We discuss possible origins of such anomalous dimensional crossovers with hole doping, 
including an interpretation based on the possible existence of a hidden quantum critical point 
near the optimally doped region. 

\end{abstract}

\pacs{74.25.Nf, 74.40.+k, 74.72.Dn, 74.78.Bz}

\keywords{high-{\it T}$_c$ cuprate, critical fluctuation, dynamic scaling, quantum criticality, dimensional crossover}

\maketitle

\section{\label{sec:Intro}Introduction}
One of the most striking features in the cuprate superconductors is 
the strong doping dependence of the critical temperature, $T_c$. 
It is well known that a plot of $T_c$ versus the carrier doping forms a bell-shaped phase diagram. 
In spite of a large number of experimental and theoretical studies on this issue, 
the physical origin of the phase diagram 
including the pseudogap phenomena above $T_c$ in the underdoped region is still debated. 

When we focus on a role of quantum criticality (QC) in the phase diagram of high-$T_c$ cuprates, 
most of theoretically proposed phase diagrams can be classified into the following two groups. 
In the first group, two quantum critical points (QCPs) are recognized 
at both endpoints of the superconducting transition line. 
However, the bell-shaped superconducting dome or anomalous properties in the pseudogap regime 
is explained by a different concept from the QC, 
for instance, the gauge field fluctuations in the $t-J$ model,~\cite{Suzumura1988,PALee1992}
or the classical phase fluctuations of superconducting orders, 
which are assumed to survive even in the pseudogap regime.~\cite{EK95}

On the other hand, in the second group, 
in addition to the two well established QCPs in the first group, 
another QCP is assumed to exist (or to be hidden) inside the superconducting dome. 
In other words, the existence of a hidden order is assumed, 
which may compete or cooperate with the superconducting order. 
In this view point, superconductivity and the anomalous behavior in the pseudogap regime 
are related to the hidden order and its large quantum fluctuation around the hidden QCP.
Various candidates for this hidden order have been proposed; 
for instance, a magnetic N\'{e}el order,~\cite{Moriya,Pines} 
an incomensurate charge-density-wave,~\cite{Castellani1995} 
a time-reversal-violating state,~\cite{Varma1997} 
a $d$-density wave,~\cite{Chakravarty2001} 
and a charge/spin stripe order.~\cite{Zaanen,KivelsonRMP2003} 
Possible candidates for such an unknown order have also been investigated 
by a more general consideration using a group-theoretic classification 
for competing orders.~\cite{Vojta2000} 
Several experiments have suggested a possibility that there is the hidden QCP 
in the vicinity of the optimally doped 
concentration.~\cite{Aeppli1997,Valla1999,Tallon1999,Balakirev2003,VDMarel2003,Ando2004} 
However, the interpretation of such experimental results is still controversial, 
and the relationship with a mechanism of superconductivity also remains unresolved. 

These two groups contrast strikingly with each other in term of 
the possibility of a hidden quantum phase transition (QPT). 
If a typical frequency scale of such a QPT was sufficiently larger than 
a thermal energy, it is difficult to disentangle quantum and thermal 
effects.~\cite{SachdevTextbook} 
This means that the classical critical dynamics can be affected by the 
quantum-fluctuations effects near the hidden QCP. 
Therefore, one can know which group is more appropriate for describing 
the phase diagram of high-$T_c$ cuprates, by investigating the 
superconducting to the normal-state transition with a wide range 
of carrier doping. 
Fortunately, the critical fluctuation effects, which reflect the transition nature, 
can be explored in the high-$T_c$ cuprates, 
because the short coherence length, the small superfluid density and the quasi-two-dimensionality 
largely enhance the classical fluctuations of the superconducting order.~\cite{ref:FFH}

In our previous paper,~\cite{KitanoPRB2006}
we reported that 
the critical charge dynamics for underdoped La$_{2-x}$Sr$_x$CuO$_4$ (LSCO) from $x$=0.07 to 0.14 was
successfully expressed by the 2D-$XY$ universality class, 
that is, Berezinskii-Kosterlitz-Thouless (BKT) picture.~\cite{KT} 
On the other hand, the critical behaviors for almost optimally doped LSCO ($x$=0.16) 
were found to be expressed by the 3D-$XY$ universality class. 
This sudden change of the universality class was highly unexpected,
because the classical critical behavior must be universal,
irrespective to the microscopic details, such as a carrier concentration.~\cite{ChakinTextbook} 

These anomalous results supported the idea that the systematic 
measurements of the critical charge dynamics as functions of 
carrier concentration are quite effective to study the phase 
diagram of high-$T_c$ cuprates. 
However, the previous work covered a part of the whole phase diagram.
Thus, it is crucially important to complete the investigation including the overdoped region.

Based on these backgrounds, 
we have investigated the critical charge dynamics of the superconducting 
to the normal-state transition for high-quality LSCO thin films 
with a wide range of the Sr concentration.
We measured the frequency dependence of the complex microwave conductivity, 
$\sigma(\omega)=\sigma_1(\omega)-i\sigma_2(\omega)$, 
which is enhanced by the superconducting fluctuation, under zero and small finite magnetic fields. 
In this paper, 
we present new results of the critical charge dynamics for optimally doped LSCO ($x$=0.15 and 0.16) 
and overdoped LSCO (from $x$=0.17 to 0.20). 
The most important finding was 
that there is another dimensional crossover from 3D to 2D between $x$=0.16 and $x$=0.18. 
To clarify the details of this anomalous behavior, 
the finite size effects were also investigated, 
by changing the film thickness or by applying small finite magnetic fields. 
Together with our previous study,~\cite{KitanoPRB2006} 
we discuss a whole picture of the critical charge dynamics 
in the phase diagram of LSCO in terms of various proposed models. 
We consider that the newly found second sudden crossover needs to be explained 
by assuming the additional hidden QCP near the optimal doping.

This paper is organized as follows. 
In Sec. II, 
we summarize the practical merits of a dynamic scaling analysis 
using the frequency-dependent complex conductivity, 
comparing with the scaling analyses using other thermodynamic and transport properties. 
In Sec. III, 
we briefly describe the preparation of high-quality LSCO thin films 
and the broadband technique to obtain the frequency-dependent microwave conductivity. 
Section IV contains all the experimental results 
on the dependence of the critical charge dynamics on the hole-doping, 
the film thickness and the magnetic field. 
In Sec. V, various proposed models are systematically discussed 
to explain the sets of our data. Finally, we conclude this work in Sec. VI.

\section{\label{sec:theory}
Dynamic scaling analysis of complex fluctuation conductivity}

The critical behaviors are characterized by the divergence of typical length and time scales 
in the vicinity of a critical point. 
In the case of classical phase transition,~\cite{ChakinTextbook} 
a correlation length, $\xi$, is often used as the typical length scale. 
$\xi$ diverges as $\xi=\xi_0|T/T_c-1|^{-\nu}$, where $\xi_0$ is the correlation length at $T=0$ 
and $\nu$ is a static critical exponent. 
On the other hand, the divergence of a correlation time, $\tau$, 
is given by $\tau=\xi^z$. Here, $z$ is a dynamic critical exponent. 
The divergence of $\xi$ and $\tau$ in quantum phase transition 
can also be defined in a similar manner,
except that a singular point is not a $T_c$ but a QCP at $T=0$. 

Fisher, Fisher, and Huse (FFH)~\cite{ref:FFH} provided 
a general formulation of the dynamic scaling hypothesis 
for the frequency-dependent complex fluctuation conductivity, $\sigma_{\rm fl}(\omega)$, 
near the superconducting transition, 
as follows, 
\begin{equation}
\sigma_{\rm fl}(\omega)\approx\xi^{z+2-d} S(\omega\xi^z). 
\label{eq:FFH1}
\end{equation}
\noindent
Here, $S(x)$ is a complex universal scaling function and $d$ is a spatial dimension. 
As was emphasized in our previous paper,~\cite{KitanoPRB2006} 
the most essential part of our analyses is that we can check the validity of this hypothesis, 
by measuring the frequency dependence of $\sigma_{\rm fl}(\omega)$. 
Following a pioneering work by Booth {\it et al.},~\cite{Booth96} 
we use both the magnitude, $|\sigma|$, 
and the phase, $\phi_\sigma(\equiv\tan^{-1}[\sigma_2^{\rm fl}/\sigma_1^{\rm fl}])$, 
of $\sigma_{\rm fl}(\omega)$ 
as scaled quantities in the scaling analysis of $\sigma_{\rm fl}(\omega)$. 
First of all, 
the phase part of $\sigma_{\rm fl}(\omega)$ at different temperatures is scaled by 
using a normalizing factor, $\omega_0$, 
along the $\omega$-direction in the plots of $\phi_\sigma$ versus $\omega$. 
Next, 
the magnitude part of $\sigma_{\rm fl}$ as a function of the normalized frequency, $\omega/\omega_0$, 
is also scaled by using another normalizing factor, $\sigma_0$, along the $|\sigma|$-direction. 
Note that $\omega_0$ and $\sigma_0$ are independently obtained in our procedure, 
since we measure data sets of $\sigma_{\rm fl}(\omega)$ as complex quantities 
with two independent components. 

If the data sets of $\phi_\sigma$ and those of $|\sigma|/\sigma_0$ 
collapse on to a single universal function of $\omega/\omega_0$, respectively, 
we find that the dynamic scaling hypothesis is satisfied. 
Thus, Eq.~(\ref{eq:FFH1}) tells us that the two scaling parameters, $\omega_0$ and $\sigma_0$, 
can be connected with the diverging length scale, $\xi$, as follows, 
\begin{eqnarray}
\omega_0&\propto&\xi^{-z} \propto |T/T_c-1|^{\nu z}, \\
\sigma_0&\propto&\xi^{z+2-d} \propto |T/T_c-1|^{-\nu (z+2-d)}.
\label{eq:FFH2}
\end{eqnarray}
By using these relationships, 
we can directly determine the two critical exponents, $\nu$ and $z$, and the dimensionality, $d$, 
through the temperature dependence of $\omega_0$ and $\sigma_0$. 
On the other hand, 
if the data sets of $\phi_\sigma$ or those of $|\sigma|/\sigma_0$ do not collapse on to a single curve, 
we find that the dynamic scaling hypothesis is broken down, 
so that we cannot extract the information of the universality class from Eqs.~(2) and (3). 
Indeed, we have encountered such a breakdown of the data collapse in $\phi_\sigma$ 
for the conventional NbN superconducting thick film, 
showing a dimensional crossover 
due to the divergence of $\xi$ beyond the film thickness.~\cite{OhashiPRB2006} 
We also observed that a broad distribution of $T_c$ caused a breakdown of the dynamic scaling, 
which will be described in Sec. V.

The dynamic scaling procedures described above yield more reliable information 
than other scaling procedures, because of the following reasons.
First of all, 
the scaling procedure of $\phi_\sigma(\omega)$ was performed in a linear scale.
This required a more strict criterion in data collapse 
than ordinary scaling procedures performed in the log-log plots, 
which easily lead to spurious scaled behaviors. 
In addition, the dynamic scaling analyses can be performed by using only the experimental data. 
In other words, we can check whether the data obey the scaling hypothesis or not, 
without any assumption. 
This is in contrast with other scaling procedures. 
For example, in the case of the scaling analysis of the $I$-$V$ 
characteristics,~\cite{I-Vcurve} the dimensionality, $d$, has to 
be assumed first. The value of $z$ is obtained from the limiting 
behavior of $V(\propto I^{(z+1)/(d-1)})$ at $T_c$. 
Only after those procedures, the data can be scaled in the plots of 
$(I/T)|T-T_c|^{-\nu(d-1)}$ versus $(V/I)|T-T_c|^{-\nu(2+z-d)}$. 
The value of $\nu$ is obtained from a successfully scaled behavior 
of the plots. 
However, the ambiguity of $T_c$ easily gives rise to large error bars 
on the obtained results of $z$ and $\nu$, which degrades 
the reliability of the scaling analysis. 
In addition, one cannot detect the breakdown of the dynamic scaling 
hypothesis due to the dimensional crossover in these procedures. 
Other scaling analyses require 
the assumpition of a particular model in advance to perform the scaling analyses as well, 
such as, the scaling analysis of the dc magnetization,~\cite{magnetization} 
the specific heat,~\cite{specific_heat} 
the thermal expansion,~\cite{thermal_ex} 
and the ac conductivity at a fixed frequency.~\cite{ac_conductivity}

Unfortunately, there was no consensus among the previous results 
on the superconducting fluctuations of high-$T_c$ cuprates, 
mostly based on the measurements described above.
We believe that such controversial results were attributed to the lack of the experimental check 
on the applicability of the static or dynamic scaling hypothesis. 
In this sense, only the dynamic scaling analysis of $\sigma_{\rm fl}(\omega)$ is exceptional, 
since no assumption is needed to check whether the data sets of $\phi_\sigma$ or $|\sigma|$ 
collapse on to a single curve or not. 
This is the reason why we adopted this method to investigate the critical charge dynamics 
near the superconducting transition of high-$T_c$ cuprates as a function of carrier doping. 

More precisely speaking, the determination of $\nu$ and $z$ is affected by the ambiguity of $T_c$, 
even if the dynamic scaling form predicted by FFH is confirmed to be satisfied. 
However, the determination of the dimensionality, $d$, is free from the ambiguity of $T_c$, 
since a product of $\omega_0$ and $\sigma_0$ is 
proportional to $\xi^{2-d}$, as derived from Eqs.~(2) and (3). 
This suggests that $\omega_0\sigma_0$ is independent of $T$ for $d$=2, 
while it increases with increasing $T$ for $d$=3, since $\nu$ is always positive. 
In addition, Eq.~(\ref{eq:FFH1}) shows us that the phase angle 
and the magnitude of $\sigma_{\rm fl}(\omega)$ behave as, 
\begin{eqnarray}
\phi_{\sigma} &\to& \frac{\pi}{2}\frac{z+2-d}{z},\\
\frac{|\sigma|}{\sigma_0} &\to& \omega^{-(z+2-d)/z},
\label{eq:FFH3}
\end{eqnarray}
respectively, as $T$ approaches $T_c$. 
This limiting behavior is also very useful to discuss the dimensionality. 
If $d$=2, then $\phi_\sigma\sim\pi/2$ and $|\sigma|\propto 1/\omega$ for $T\to T_c$, 
independent of $z$. 
On the other hand, for $d$=3, it is expected that 
$\phi_\sigma\sim\pi/4$ and $|\sigma|\propto 1/\omega^2$ for $T\to T_c$, 
since a fully relaxational dynamics with $z$$\approx$2 
(that is, the so-called model-$A$ dynamics 
in the Hohenberg and Halperin classification~\cite{HHreview}) 
is expected for a charged superfluid.~\cite{ref:FFH} 
Thus, we can discuss the dimensionality of the superconducting phase transition 
through the limiting behavior of $\phi_\sigma$ and $\omega_0\sigma_0$ in the vicinity of $T_c$, 
even if the precise value of $T_c$ is unknown. 
This is particularly important when we discuss the dimensionality as a function of carrier doping.

\section{\label{sec:3}Experimental}

\subsection{\label{sec:3-1}High quality LSCO films}
\begin{table}
\caption{
\label{tab:sample}
Various parameters of LSCO films. See the text for details.}
\begin{ruledtabular}
\begin{tabular}{cccccc}
$x$ & $t$ (nm) & $\rho$ (m$\Omega$cm) & $T_c^R$ (K) & $\Delta T_c (K)$ & $T_c^{\rm scale}$ (K)\\
\hline
0.07 & 460 & 0.78 & 18.0 & 3.6 & 18.2 \\
0.10 & 220 & 0.35 & 28.2 & 1.7 & 28.5 \\
0.12 & 230 & 0.28 & 33.7 & 1.5 & 33.7 \\
0.14 & 270 & 0.19 & 38.9 & 0.5 & 38.9 \\
0.15 & 150 & 0.22 & 31.6 & 1.8 & 32.1 \\
0.16 & 140 & 0.12 & 35.5 & 1.6 & 35.9 \\
0.17 & 140 & 0.15 & 34.4 & 1.7 & 34.7 \\
0.17 & 115 & 0.14 & 34.4 & 1.2 & 34.7 \\
0.18 & 240 & 0.11 & 34.5 & 1.3 & 34.8 \\
0.18 & 120 & 0.15 & 27.9 & 2.7 & -    \\
0.18 &  60 & 0.15 & 27.8 & 1.5 & 28.2 \\
0.19 & 135 & 0.09 & 26.7 & 1.0 & 26.9 \\
0.20 &  90 & 0.05 & 32.4 & 1.7 & 32.8 \\
\end{tabular}
\end{ruledtabular}
\end{table}

Epitaxial LSCO thin films covering a wide range from $x$=0.07 to 0.20 were 
grown on LaSrAlO$_4$ (001) substrates by a pulsed laser deposition technique. 
Details of the growth condition were described elsewhere.~\cite{Tsukada06,Tsukada04,Lavrov03} 
Pure ozone in an atmosphere of 10~mPa was used for oxidation during 
both the deposition and postannealing processes in the growth chamber, 
in order to obtain superconducting LSCO films with very low residual resistivity. 
The excellent $c$-axis orientation of each film was confirmed by the X-ray diffraction. 
The rocking curve of the 002 reflection showed the full width at half maximum (FWHM) 
of about 0.2$^\circ$, which was almost the same as FWHM of LaSrAlO$_4$ (LSAO) substrate, 
suggesting that the mosaicness of the film 
is not so much different from that of the substrate.~\cite{Tsukada04}

All films were carefully annealed in air (or O$_2$ gas for $x$=0.20) at 700~$^\circ$C for 2-4 hours, 
following the heat treatment procedure developed for high-quality LSCO single crystals.~\cite{Komiya02} 
The annealed samples were rapidly quenched down to low temperatures using liquid nitrogen. 

\begin{figure}
\includegraphics[width=0.7\linewidth]{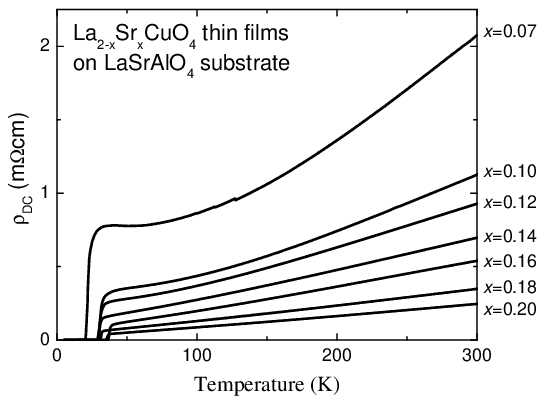}
\caption{
Temperature dependence of the dc resistivity for LSCO thin films with $x$=0.07 to 0.20.
}
\label{fig:dc}
\end{figure}

The in-plane dc resistivity, $\rho_{\rm dc}$, of each film 
was measured using a standard four-probe method. 
Table I lists a thickness, the resistivity at $T$=50~K, $T_c$ defined by zero resistance, 
$\Delta T_c$ defined by the 10\%-90\% transition width, 
and $T_c$ used in the dynamic scaling analysis for all the samples. 
As shown in Fig.~\ref{fig:dc} and Table I, 
we confirmed that the absolute value of $\rho_{\rm dc}$ just above $T_c$ 
systematically decreased with increasing hole doping within the experimental error, 
and also that it agreed with the reported best values for LSCO thin films~\cite{Sato97,Bozovic02} 
and single crystals~\cite{Ando04} with the same value of $x$, within the factor of 2. 
Together with the narrow rocking curve observed by the x-ray diffraction, 
these properties show that the thin films under this study are of so high quality 
that the critical phenomenon near $T_c$ deserved to be investigated experimentally. 
Note that the transitions appear to be broad, especially for underdoped films. 
However, this is a consequence of the fluctuation, and is not due to the inhomogeneity of $T_c$.
The only exception is the 120~nm thick film with $x$=0.18, which will be referred in Sec. V 
as an example of the breakdown of the dynamics scaling due to the inhomogeneity of $T_c$.

As was already described in our previous paper,~\cite{KitanoPRB2006} 
for the purpose of investigating the critical charge dynamics in a whole region of the phase diagram, 
LSCO is an ideal system with a simple layered structure, 
where the hole concentration can be widely controlled. 
In addition, the use of LSAO substrate for LSCO films provides an advantage over bulk crystals of LSCO, 
because the tetragonal symmetry of LSAO substrate prevents the corrugation in the CuO$_2$ plane, 
which is attributed to a staggered rotation of CuO$_6$ octahedra 
in low-temperature orthorhombic (LTO) phase of bulk LSCO.~\cite{Tsukada04} 
Thus, we can expect that ideal flat square CuO$_2$ planes, 
which are free from disorders due to corrugations and twin boundaries, 
are realized in LSCO films on LSAO substrate.

\subsection{\label{sec:3-2}Microwave broadband technique}

When the film thickness, $t$, is sufficiently smaller than the skin depth, $\delta$, the frequency-dependent microwave complex conductivity, $\sigma(\omega)$, can be obtained from the complex reflection coefficient, $S_{11}(\omega)$, as follows,~\cite{KitanoPhysicaC} 
\begin{equation}
\sigma(\omega)=\frac{1}{tZ_0}\frac{1-S_{11}(\omega)}{1+S_{11}(\omega)},
\label{eq:sigma}
\end{equation}
where $t$ and $Z_0(=377~\Omega$) are the film thickness and the impedance of free space, respectively. 

We measured the frequency dependence of $S_{11}(\omega)$ 
in the frequency range between 45~MHz and 10~GHz, 
by using a vector network analyzer (HP 8510C) in a step-sweep mode. 
The thin film sample with the electrodes in a Corbino disk shape 
was terminated to the end of a coaxial cable through a modified jack-to-jack coaxial adaptor. 
A stable electrical contact between the sample and the modified coaxial adaptor 
was realized by two springs, following the work by Booth {\it et al.}~\cite{BoothRSI} 
One spring was embedded into a center conductor pin of the adaptor, 
and the other spring was located at the backside of the LSAO substrate
Other details of the experimental setup will be described elsewhere.~\cite{KitanoRSI} 

\begin{figure}
\includegraphics[width=0.7\linewidth]{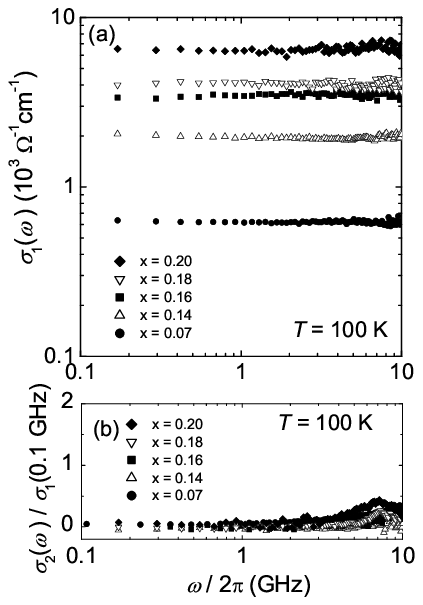}
\caption{
(a) Frequency dependence of $\sigma_1$ for LSCO thin films with $x$=0.07 to 0.20 at $T$=100~K. 
(b) Frequency dependence of $\sigma_2$ normalized by $\sigma_1$ at $0.1$~GHz for the same films. 
For all films, $\sigma_1$ is independent of frequency and $\sigma_2$ is much smaller than $\sigma_1$, 
suggesting that the charge dynamics in the normal state 
can be safely regarded as in the Hagen-Rubens limit of the Drude conductivity.
} 
\label{fig:T100K}
\end{figure}

In practice, the measured reflection coefficient, $S_{11}^{\rm meas}(\omega)$, 
is affected by the attenuation and the phase shifts in the intervening coaxial cable, 
giving rise to large systematic errors at lower temperatures and higher frequencies. 
Thus, $S_{11}^{\rm meas}(\omega)$ is given by the following equation,~\cite{Pozar} 
\begin{equation}
S_{11}^{\rm meas}(\omega)=E_D(\omega)+\frac{E_R(\omega)S_{11}(\omega)}{1-E_S(\omega)S_{11}(\omega)}, 
\label{eq:error}
\end{equation}
where, $E_D(\omega)$, $E_R(\omega)$, and $E_S(\omega)$ are complex error coefficients, 
representing the directivity, the reflection tracking, and the source mismatching, respectively. 
A standard method to determine these unknown error coefficients 
is to measure three known reference samples as a function of frequency at each temperature. 
Typically, we used a gold film as a short standard, a NiCr film as a load standard, 
and a teflon sheet as an open standard, following the work by Stutzman {\it et al}.~\cite{StutzmanRSI} 
By performing this standard calibration procedure, 
both $\sigma_1(\omega)$ and $\sigma_2(\omega)$ for all LSCO films 
were successfully obtained in the frequency range at least between 0.1~GHz and 10~GHz, 
as shown in Fig.~\ref{fig:T100K}. 
For all films, we confirmed that the skin depth $\delta$ at 10~GHz, 
which was given by $\sigma(\omega)$, was sufficiently longer than the film thickness $t$ 
over the whole measured temperature range including the vicinity of $T_c$, 
and that the frequency dependence of both $\sigma_1(\omega)$ and $\sigma_2(\omega)$ well above $T_c$ 
could be regarded as those in the Hagen-Rubens limit of the Drude conductivity.~\cite{KitanoPhysicaC} 

\subsection{\label{sec:3-3}Calibration procedure to obtain complex fluctuation conductivity}

The standard calibration procedure using three known standards 
is based on the high reproducibility between four measurements 
including three known reference samples and an unknown LSCO sample. 
In our measurement apparatus, 
the measured $S_{11}$ was reproduced typically within 0.05~dB in the magnitude 
and 0.2$^\circ$ in the phase of $S_{11}$ at 1~GHz.
As shown in Fig.~2, this reproducibility was enough for us to obtain $\sigma(\omega)$ 
between 0.1~GHz and 10~GHz for LSCO films far above $T_c$. 
However, as was already reported by our previous paper,~\cite{KitanoPRB2006} 
we found that the present reproducibility 
was insufficient to obtain $\sigma(\omega)$ in the vicinity of $T_c$. 
This is because a small unexpected difference 
between the phase of $S_{11}^{\rm  short}$ and that of $S_{11}^{\rm load}$ 
gave rise to a large error in $\sigma(\omega)$ in the vicinity of $T_c$, 
as suggested by Eq.~(\ref{eq:sigma}). 

In order to overcome this difficulty, 
we modified the calibration procedure as follows. 
First of all, we regarded both the magnitude and phase parts of $S_{11}(\omega)$ for the LSCO sample, 
which were measured at a temperature ($T=T_0$) well above $T_c$, as those of $S_{11}^{\rm load}$, 
based on the observation that the normal-state conductivity of the LSCO films was regarded as 
that in the Hagen-Rubens limit of the Drude conductivity. 
Next, we equated the phase of $S_{11}^{\rm short}$ to that of $S_{11}^{\rm load}$ at $T=T_0$, 
based on the assumption that a small difference between them was purely due to the experimental error. 
Finally, we assumed that the phase parts of both $S_{11}^{\rm short}$ and $S_{11}^{\rm load}$ 
were independent of temperature in a narrow range between $T_c$ and $T_0$. 
Thus, the phase of $S_{11}^{\rm short}$ agreed with that of $S_{11}^{\rm load}$ 
at any temperature from $T_c$ to $T_0$, as expected for the ideal load and short standards. 
We also confirmed the validity of this procedure 
by the same measurements of conventional NbN superconducting thin films 
as a reference.~\cite{OhashiPRB2006} 

If $\sigma(\omega)$ was obtained at each temperature in the vicinity of $T_c$ 
through the modified calibration procedure, 
the complex fluctuation conductivity, $\sigma_{\rm fl}(\omega,T)$, is given by the following equation, 
\begin{equation}
\sigma_{\rm fl}(\omega,T) \equiv \sigma(\omega,T)-\sigma_n(T).
\label{eq:sigma_fl}
\end{equation}
Here, $\sigma_n(T)$ is the normal-state component included in $\sigma(\omega,T)$, 
which agrees with the normal-state conductivity far above $T_c$. 
Although the change of $\sigma_n(T)$ with decreasing temperature 
was estimated by the linear extrapolation of dc conductivity from higher temperatures, 
we found that the dynamic scaling behavior of the extracted $\sigma_{\rm fl}(\omega)$ 
in the vicinity of $T_c$ was insensitive to the temperature dependence of $\sigma_n(T)$.

\section{\label{sec:results}Results}
\subsection{\label{sec:4-1}Doping dependence in zero magnetic field}

\begin{figure}
\includegraphics[width=\linewidth]{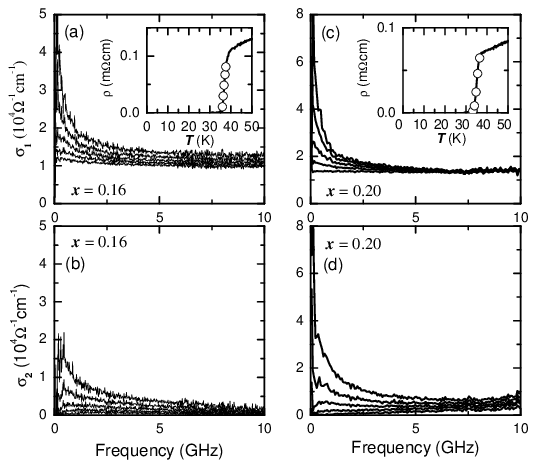}
\caption{
Frequency dependence of (a) $\sigma_1$ and (b) $\sigma_2$ of the $x$=0.16 film at $T$=36.0-37.6~K, 
and (c) $\sigma_1$ and (d) $\sigma_2$ for the $x$=0.20 film at $T$=33.2-36.5~K.
All temperatures are just above $T_c$. 
Both $\sigma_1$ and $\sigma_2$ showed a remarkable enhancement in the low frequency limit 
as temperature approaches $T_c$. 
The insets show the temperature dependence of dc resistivity. 
Each circle represents the temperature of each data in the main panels.
}
\label{fig:sigma}
\end{figure}

\begin{figure}
\includegraphics[width=\linewidth]{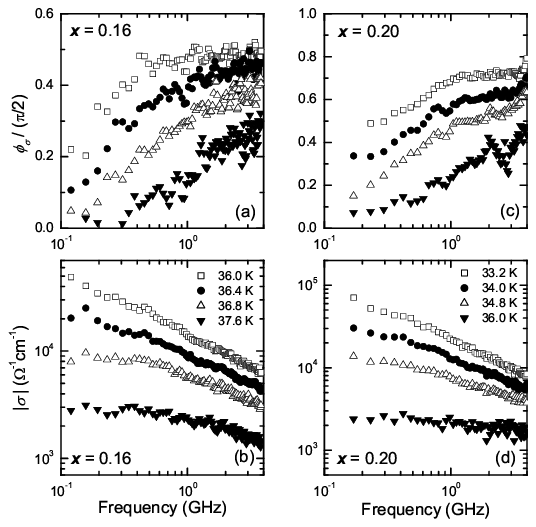}
\caption{
Frequency dependence of (a) the phase, $\phi_{\sigma}$, 
and (b) the magnitude, $|\sigma|$, of the fluctuation induced complex conductivity for the $x$=0.16 film , and 
(c) $\phi_{\sigma}$ and (d) $|\sigma|$ for the $x$=0.20 film. 
}
\label{fig:MagPhase}
\end{figure}

Figure \ref{fig:sigma} shows the typical frequency dependence of $\sigma(\omega)$ 
for the optimally doped LSCO ($x$=0.16) and the overdoped LSCO ($x$=0.20) 
at several temperatures near $T_c$ in zero magnetic field. 
We observed that both $\sigma_1(\omega)$ and $\sigma_2(\omega)$ in the low frequency limit 
diverged rapidly as the temperature approached $T_c$ from above, 
suggesting that the contribution of the superconducting fluctuations to $\sigma(\omega)$ 
was evident with decreasing temperature. 
This behavior was more clearly demonstrated 
by plotting the phase and the magnitude of $\sigma_{\rm fl}(\omega)(=\sigma(\omega)-\sigma_n)$ 
as a function of frequency. 
Figures~\ref{fig:MagPhase}(a) and (c) shows 
that the plots of the dimensionless phase angle, $\phi_\sigma$, 
of $\sigma_{\rm fl}$ versus frequency move to a lower frequency region with decreasing temperature, 
suggesting that a characteristic time scale of the superconducting fluctuations 
is critically slowing down. 
On the other hand, as shown in Figs.~\ref{fig:MagPhase}(b) and (d), 
the magnitude, $|\sigma|$, of $\sigma_{\rm fl}$ increases rapidly with decreasing temperature, 
suggesting a critical divergence of the superconducting fluctuations. 

\begin{figure*}
\includegraphics[width=\linewidth]{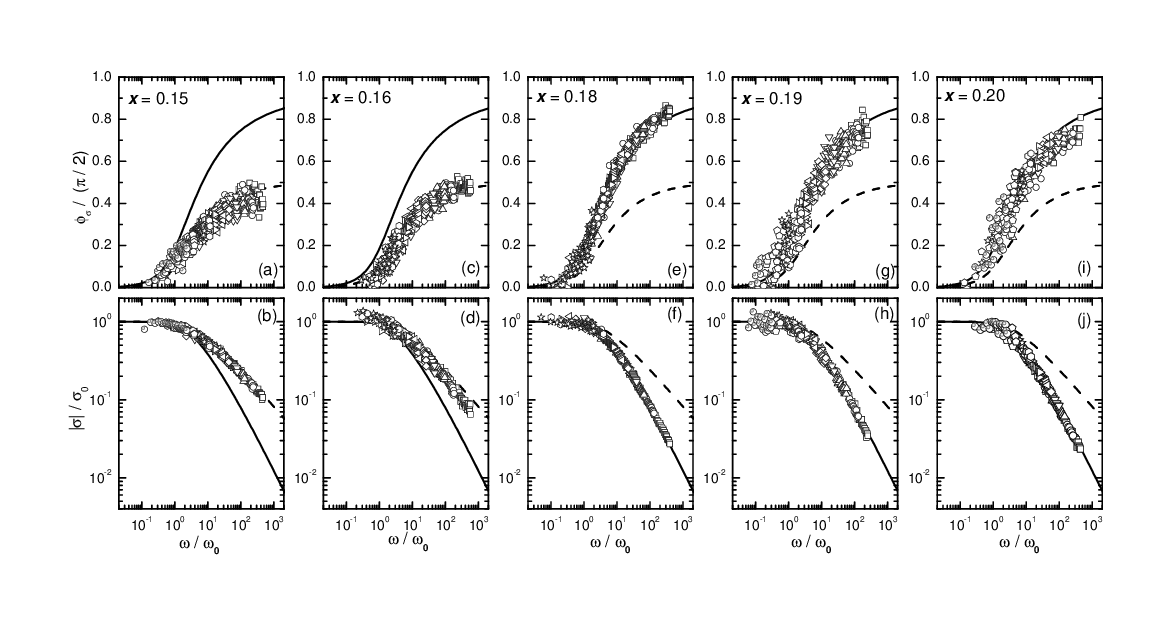}
\caption{\label{fig:scaling}
Scaled data of the phase (upper panels) and the magnitude (lower panes) 
of $\sigma_{\rm fl(\omega)}$ for the LSCO films with various carrier concentrations. 
The phase and the magnitude is normalized by $\pi/2$ 
and the scaling parameter, $\sigma_0$, respectively.
The temperature range of the scaled data for each film is from $T$=32.4~K to 35.0~K ($x$=0.15), 
from $T$=36.0~K to 37.6~K ($x$=0.16), from $T$=28.2~K to 31.0~K ($x$=0.18), 
from $T$=27.0~K to 29.0~K ($x$=0.19), and from $T$=33.2 to 36.5~K ($x$=0.20), respectively.
Solid (dashed) lines are the 2D (3D) Gaussian scaling functions.
}
\end{figure*}

Following the scaling procedures described in Sec.~\ref{sec:theory}, 
we tested whether these behaviors were truly attributed to the critical fluctuations 
of the superconducting order or not. 
Figure \ref{fig:scaling} shows the results of the data scaling 
of $\phi_\sigma$ (upper panels) and $|\sigma_{\rm fl}|$ (lower panels) for 
the LSCO films with five Sr concentrations from $x$=0.15 to $x$=0.20. 
As clearly indicated in Fig.~\ref{fig:scaling}, 
we confirmed that both of $\phi_\sigma$ and $|\sigma|$ successfully 
collapsed on to single curves for all Sr concentrations, 
strongly suggesting that the dynamic scaling hypothesis was successfully satisfied 
so that we could investigate the change of the critical behavior 
as a function of the Sr concentration along this line. 

\begin{figure}
\includegraphics[width=0.9\linewidth]{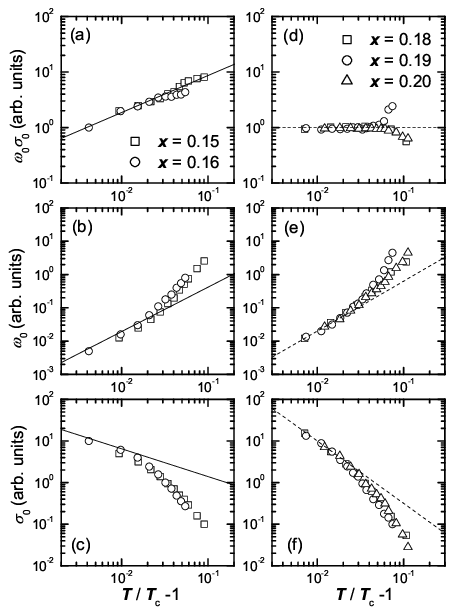}
\caption{
Temperature dependence of the obtained scaling parameters, $\omega_0\sigma_0$ (top), $\omega_0$ (middle), and $\sigma_0$ (bottom). 
Left (Right) panels are for the optimally doped (the overdoped) LSCO. 
Solid lines in the left panels are calculations given by Eqs.~(2) and (3) with $\nu$=0.67, $z$=2, $d$=3, suggesting the 3D-$XY$ critical charge dynamics. 
Dashed lines in the right panels are similar calculations with $d$=2 and $\nu z$=1.5.}
\label{fig:para}
\end{figure}

The solid and dashed lines depicted in Fig.~\ref{fig:scaling} 
were the 2D and 3D Gaussian scaling functions, respectively, 
which were calculated by Schmidt.~\cite{Schmidt} 
They play roles of the guide to eyes for the dimensionality in the critical charge dynamics, 
suggested by Eqs.~(4) and (5). 
We found that the optimally doped LSCO ($x$=0.15, 0.16) showed the 3D-like critical behavior 
while the overdoped LSCO ($x$=0.18$\sim$0.20) showed the 2D-like critical behavior.~\cite{OhashiM2S} 
This suggestion was quantitatively confirmed by the plots of the scaling parameters, 
$\omega_0$ and $\sigma_0$, versus the reduced temperature, $T/T_c-1$, as shown in Fig.~\ref{fig:para}. 
Here, we used the values of $T_c^{\rm scale}$ as $T_c$. 
In the dynamic scaling theory by FFH,~\cite{ref:FFH} the plots of $\omega_0\sigma_0(\propto\xi^{2-d})$ 
are expected to be independent of $T/T_c-1$ for $d$=2, 
while they are expected to increase monotonously with $T/T_c-1$ for $d$=3. 
Thus, the plots of $\omega_0\sigma_0$ in Figs.~\ref{fig:para}(a) and (d) clearly indicate 
that the dimensional crossover from 3D to 2D occurred near the boundary 
between the optimally doped region and the overdoped region. 

We also confirmed that $\omega_0$ and $\sigma_0$ for the optimally doped LSCO 
with $x$=0.15 and $x$=0.16 agreed with those for 
the relaxational 3D-$XY$ universality class with $\nu$=0.67 and $z$$\approx$2,~\cite{Wickham00} 
as shown in Figs.~\ref{fig:para}(b) and (c). 
On the other hand, $\omega_0$ and $\sigma_0$ for the overdoped LSCO with $x$=0.18-0.20 
were found not to show an exponential singularity expected for the 2D-$XY$ universality class, 
but rather to show a divergence with an exponent of $\nu z$$\sim$1.5, 
as shown in Figs.~\ref{fig:para}(e) and (f). 
This shows a sharp contrast to the behavior for the underdoped LSCO with $x$=0.07-0.14, 
as already reported in the previous paper.~\cite{KitanoPRB2006} 

We note that the critical behavior with $\nu z$$\sim$1.5 observed for the overdoped LSCO 
is different from both the BKT transition classified in the 2D-$XY$ universality class 
and the 2D Gaussian fluctuation whose critical exponents are $\nu$=0.5 and $z$=2.
Thus, we refer this unusual behavior to the 2D-``U", as an unknown universality class.
In the next section, we discuss the origin of this unusual behavior, 
in terms of various possible candidates. 

Figure \ref{fig:phase1} shows a summary of our results 
about the critical charge dynamics of LSCO as a function of the Sr concentration. 
The temperature region where the suggested universality class became dominant 
was determined by fitting of the temperature dependence of $\omega_0$ to corresponding models. 
Together with our previous study for the underdoped LSCO,~\cite{KitanoPRB2006} 
we found that two kinds of the sudden dimensional crossovers took place across the phase diagram. 
One is the dimensional crossover from 2D-$XY$ to 3D-$XY$ between $x$=0.14 and $x$=0.15, 
while the other is the dimensional crossover from 3D-$XY$ to 2D-``U" between $x$=0.16 and $x$=0.18. 
The first dimensional crossover was already reported in our previous study,~\cite{KitanoPRB2006} 
while the second was observed by this work for the first time. 

In the following two subsections, we first investigate the critical behavior in the overdoped LSCO 
in more detail, in terms of finite size effects. 

\subsection{\label{sec:thick}Finite size effect} 

In this study, we used relatively thin LSCO films in the overdoped regime, as listed in Table I. 
This is because, for the reliable data of $\sigma_{\rm fl}(\omega)$ 
to be obtained by the microwave broadband technique, 
the sheet resistance just above $T_c$ is needed to be larger than about 2~$\Omega$. 
This condition requires us to use thinner films for less resistive overdoped LSCO.
However, the use of thinner films leads to a suspicion 
that the observed 2D-like critical behavior for the overdoped LSCO 
is caused by a finite film thickness. 
Indeed, we have demonstrated that the dynamic fluctuations in the superconducting NbN thin films 
showed the 2D Gaussian fluctuations, 
which were introduced by the divergence of $\xi$ beyond the film thickness.~\cite{OhashiPRB2006} 

In order to investigate this problem, 
we measured the films with different thickness near the Sr concentration 
where the second dimensional crossover occurred. 
Figures \ref{fig:scaling17t} and \ref{fig:scaling18t} 
show the scaled data of $\phi_\sigma$ for $x$=0.17 ($t$=115, 140~nm) 
and $x$=0.18 ($t$=60, 240~nm), respectively. 

First of all, as shown in Fig.~\ref{fig:scaling17t}, 
the behavior of $\phi_\sigma(\omega)$ scaled by $\omega_0$ for $x$=$0.17$ 
showed the 2D-like behavior for the thicker film ($t$=140~nm), 
and the 3D-like behavior for the thinner film ($t$=115~nm).
It is clear that this change of the dimensionality cannot be explained by the finite size effect, 
in contrast to the previous results on the NbN films.~\cite{OhashiPRB2006} 
Rather, this result implied a possibility that 
the intrinsic dimensional crossover occurs suddenly near $x$=0.17.

The behaviors of $\phi_\sigma$ for $x$=0.18 agreed with the 2D-like critical charge dynamics 
even for the thicker film ($t$=240~nm), as shown in Fig. \ref{fig:scaling18t}.
By comparing this result with that of the film with $x$=0.16 ($t$=140~nm) 
which showed the 3D-$XY$ critical charge dynamics, 
it provided us the following important insight.
As shown in Fig.~\ref{fig:xic-thickness}, 
the 2D-like critical fluctuations for the thicker film could be observed 
in the temperature range from 1.006$T_c$ to 1.02$T_c$. 
If we consider that the 2D behavior is due to the finite-thickness effect 
on the 3D-$XY$ critical behavior, 
$\xi$ should be larger than the film thickness ($t$=240~nm) below 1.02$T_c$.
This requires that $\xi_0$ is larger than 20~nm.
On the other hand, the 3D-$XY$ fluctuations for $x$=0.16 was observed 
in the temperature range from 1.004$T_c$ to 1.02$T_c$. 
This means that $\xi(T)$ is not larger than $t$=140~nm above $T$=1.004$T_c$, 
requiring that $\xi_0$ is less than 3.5~nm. 
Thus, a difference of $\xi_0$ between $x$=0.16 and $x$=0.18 
need to be larger than 6 times in magnitude. 
This seems to be too large to be explained by the change of $T_c$ 
or the superconducting gap with hole doping. 
Therefore, 
the observed behaviors in $\phi_\sigma$ for these films with different thickness never supported 
that the 2D-like critical behavior near $x$=0.18 was attributed to the finite thickness effect. 
The same conclusion was also obtained from the scaled data of $|\sigma|$. 

\subsection{\label{sec:mag}Effects of external magnetic field} 
Another way to investigate a finite size effect in the critical charge dynamics 
is the use of an external magnetic field parallel to the film. 
If a magnetic field larger than the lower critical field was applied to a type-II superconductor, 
vortices can penetrate into the superconductor. 
Since the vortex pinning effect is negligibly small near $T_c$, 
vortices are distributed uniformly over the sample.
At the core of each vortex, 
the growth of the superconducting order parameter is strongly supressed, 
suggesting the divergence of $\xi$ due to the superconducting fluctuation 
is cut off by a finite distance between vortices. 
Thus, the external magnetic field can be used as a tunable probe 
to study the finite size effect, since the mean distance between vortices, 
$\ell_v$, is proportional to $\sqrt{\Phi_0/B}$. 

Of course, it is crucially important to distinguish the critical charge dynamics 
under a magnetic field from the vortex dynamics in the mixed state. 
For this purpose, we measured four LSCO films ($x$=0.07, 0.16, 0.17 and 0.18) 
by applying magnetic field perpendicular to the CuO$_2$ planes 
up to 1~T systematically and investigated the obtained results very carefully.

Figure \ref{fig:scaling07H} shows the scaled data of $\phi_\sigma$ and $|\sigma|$ 
for the underdoped LSCO ($x$=0.07) at several constant magnetic fields. 
We found that both $\phi_\sigma$ and $|\sigma|$ were scaled successfully up to $B$=0.1~T, 
while neither of them were collapsed on to single curves at $B$=1~T. 
This suggests that the contribution of the vortex dynamics is almost negligible 
up to at least $B$=0.1~T, whereas it cannot be neglected at $B$=1~T. 
As shown in Figs.~\ref{fig:para07H}(a) and (b), 
we confirmed again that the behavior of $\omega_0$ and $\sigma_0$ obtained at $B$=0~T 
excellently agreed with the BKT theory in the temperature range from $1.02T_c$ to $1.2T_c$. 
Note that $\omega_0^{-1}\propto\sigma_0\propto\xi_{\rm KT}^2$ 
in the relaxational 2D-$XY$ universality class. 
Here, the BKT correlation length, $\xi_{\rm KT}$, 
shows an exponential divergence in a critical region, as follows, 
\begin{equation}
\xi_{\rm KT}=\xi_0\exp \left[ b/\sqrt{T/T_c-1} \right], 
\label{xiKT}
\end{equation}
where, $b$ is a numerical constant, which was determined to be $\sim$ 0.2 in this study. 

The plots of the correlation length, $\xi$, given by $1/\sqrt{\omega_0}$ 
as a function of $1/\sqrt{T/T_c-1}$ 
were found to be very useful to understand the finite size effect due to the external magnetic field. 
As is shown in Fig.~\ref{fig:para07H}(c), 
the divergence of $\xi$ in the vicinity of $T_c$ was suppressed with increasing magnetic field. 
This behavior can be understood by considering two kinds of the characteristic length scale; 
the BKT correlation length, $\xi_{\rm KT}$, and the mean distance between vortices, $\ell_v$. 
In the temperature region not very close to $T_c$, where $\xi_{\rm KT}\ll\ell_v $, 
the logarithmic interaction between a vortex and an anti-vortex, 
which is essential to the BKT transition, is not screened by the externally induced vortices. 
Thus, the exponential divergence due to $\xi_{\rm KT}$ is observed. 
On the other hand, in the vicinity of $T_c$, where $\xi_{\rm KT}\geq\ell_v$, 
the externally induced vortices can screen out the interaction between vortex and anti-vortex, 
leading to a crossover of the characteristic length scale from $\xi_{\rm KT}$ to $\ell_v$. 
Thus, the growth of $\xi$ is suppressed as $\xi$ approaches $\ell_v$. 
We confirmed that the saturated value of $\xi$ in the limit of $T_c$ 
did not exceed $\ell_v$ at each magnetic field, 
by using a reasonable assumption that $\xi_0$$\sim$3~nm. 
In summary, the result of Fig.~\ref{fig:para07H} is also consistent with the BKT picture 
for the fluctuation of the $x$=0.07 sample.

Figures \ref{fig:scaling16H} and \ref{fig:para16H} show the results of dynamic scaling analyses 
for the optimally doped LSCO ($x$=0.16) at several constant magnetic fields. 
We found that the data collapse of the scaled $\phi_\sigma$ and $|\sigma|$ to single curves 
were successful up to $B$=1~T, 
suggesting that the contribution of vortex dynamics was almost negligible 
in the measured magnetic field region. 
In addition, all the results clearly indicated that the 3D-$XY$ critical charge dynamics 
was maintained up to $B$=1~T. 
Thus, the external magnetic field 
seemed not to give any change to the critical charge dynamics near $T_c$, 
in contrast to the case of the BKT transition. 
This is because the power-law like divergence of $\xi(T)$ in the 3D-$XY$ critical charge dynamics 
is much slower than the exponential divergence of $\xi_{\rm KT}(T)$. 
Indeed, by assuming that $\xi_0$$\sim$3~nm and $\nu$=0.67, 
we could estimate that $\xi(T)$ for this film ($x$=0.16) 
was less than 31~nm in the observed temperature range at $B$=1~T. 
Since $\ell_v$ is estimated to be $\sim$ 45~nm at $B$=1~T, 
we can safely conclude that $\xi(T)$ is always smaller than $\ell_v$ even at $B$=1~T, 
strongly suggesting that the critical behavior is never affected by the external magnetic field 
at least up to $B$=1~T. 

Figures \ref{fig:scaling18H} and \ref{fig:para18H} show the results of dynamic scaling analyses 
at several constant magnetic fields for the overdoped LSCO film ($x$=0.18 and $t$=60~nm). 
As was already pointed out, the critical charge dynamics of this film in zero magnetic field 
was represented by an unknown universality class in the 2D system (2D-``U"). 
Figure \ref{fig:scaling18H} clearly indicates 
that the scaled data of both $\phi_{\sigma}$ and $|\sigma|$ up to $B$=1~T 
were very similar to those at $B$=0~T. 
In addition, the temperature dependence of the scaling parameters at all magnetic fields 
except for $B$=1~T excellently agreed with each other, as shown in Fig.~\ref{fig:para18H}. 
These results strongly suggest that the small applied magnetic field (at least up to 0.3~T) 
did not affect the critical charge dynamics of the overdoped LSCO.

Thus, for both $x$=0.16 and $x$=0.18, 
the magnetic field dependence of the critical temperature, $T_c(B)$, 
which was experimentally determined through the dynamic scaling analysis at finite magnetic field 
below $B$=0.3~T, 
can be regarded as the same as the temperature dependence of the upper critical field, $B_{c2}(T)$, 
as shown in the inset of Fig.~\ref{fig:nu}. 
Note that $B_{c2}$ is not a phase transition point in practice.
There is only a first-order transition of vortices at a lower field.
However, the concept of $B_{c2}$ as the mean-field critical field still makes sense.
This provides us another route to estimate $\nu$ and $\xi_0$, 
by using the following relationship, 
\begin{equation}
B_{c2}(T)=\frac{\phi_0}{2\pi\xi(T)^2}
=\frac{\phi_0}{2\pi}\frac{(1-T/T_{c0})^{2\nu}}{\xi_0^2}, \label{eq:Bc2}
\end{equation}
where $T_{c0}$ is the critical temperature at $B$=0~T. 
As shown in the main panel of Fig.~\ref{fig:nu}, 
we confirmed that the plots of the applied magnetic field, $B$, 
as a function of $(1-T_c(B)/T_{c0})$ for the optimally doped LSCO ($x$=0.16), 
agreed with a line of $B_{c2}(T)=\phi_0/2\pi\xi(T)^2$ with $\nu$=0.67 and $\xi_0$$\sim$3~nm, 
as suggested by the dynamic scaling analyses for $x$=0.16. 
This supports the idea that the plots of $B$ as a function of $(1-T_c(B)/T_{c0})$ 
can be regarded as those of $B_{c2}(T)$ as a function of $(1-T/T_{c0})$. 

Based on this idea, 
we estimated that $\nu$$\sim$0.9 and $\xi_0$$\sim$3~nm for the overdoped LSCO ($x$=0.18). 
As shown in Fig.~\ref{fig:nu}, it is clear that 
the value of $\nu$ for $x$=0.18 is larger than the values for the GL theory ($\nu_{\rm GL}$=0.5) 
and the 3D-$XY$ universality class ($\nu_{{\rm 3D}XY}$=0.67). 
We found that $\xi(T)$ increased up to 70~nm in the measured temperature range at $B$=0.3~T, 
by using the estimated values of $\nu$ and $\xi_0$. 
Since $\ell_v$ was estimated as 83~nm at $B$=0.3~T, 
$\xi$ was always smaller than $\ell_v$ within the measured temperature range 
at least up to $B$=0.3~T, 
which confirmed that the critical charge dynamics, not the vortex dynamics, was observed.
Together with the value of $\nu z(\sim 1.5)$ obtained at $B$=0~T, 
we obtained that $z$$\sim$1.7 for $x$=0.18, 
which was slightly smaller than the well-known value for the relaxational dynamics ($z$$\approx$2). 
Thus, the experimental results for $x$=0.18 suggested that the 2D-``U" universality class 
had the critical exponents of $\nu$$\sim$0.9 and $z$$\sim$1.7. 

In contrast to the results for $x$=0.16 and 0.18, 
the behavior of $\phi_{\sigma}$ and $|\sigma|$ for $x$=0.17 (the thicker film, $t$=140~nm) 
was largely dependent on the applied magnetic field. 
Up to 0.1~T, the scaled data of $\phi_{\sigma}$ and $|\sigma|$, 
and the temperature dependences of the scaling parameters were similar to those at $B$=0~T, 
as shown in Figs.~\ref{fig:scaling17H} and \ref{fig:para17H}, respectively.
By assuming that $\xi_0$ was about 3~nm, which was the same value as those used for $x$=0.16 and 0.18, 
$\xi(T)$ at 1.02$T_c$ was estimated to be $\sim$ 100~nm for the 2D-``U" critical charge dynamics.
Since $\ell_v$ was about 140~nm at $B$=0.1~T, $\xi(T)$ did not exceed $\ell_v$ up to $B$=0.1~T.
Thus we confirmed that the contribution of vortex dynamics was negligible 
and that the 2D-``U" critical charge dynamics retained up to 0.1~T.
However, above 0.3~T, we found that the values of $\phi_\sigma(\omega/\omega_0)$ 
in the high frequency limit clearly decreased 
with increasing magnetic field, 
and also that the frequency dependence of $|\sigma(\omega/\omega_0)|$ was weakened, 
as shown in Fig.~\ref{fig:scaling17H}. 
Within the dynamic scaling theory,~\cite{ChakinTextbook,ref:FFH} 
these results seem to suggest that the dimensionality in the critical charge dynamics 
changes from 2D to 3D by applying magnetic field. 
Figure~\ref{fig:para17H} also shows that the temperature dependence of 
$\sigma_0\omega_0(\propto\xi^{2-d})$ was changed markedly to the 3D-like above 0.3~T. 
However, we could not clearly confirm the 3D-$XY$ critical charge dynamics, 
since $\omega_0$ and $\sigma_0$ could be fitted by the critical exponents 
of the 3D-$XY$ universality class only in a very narrow region between 1.01$T_c$ and 1.03$T_c$. 
This suggests a possibility that another contribution affects the critical behavior above 0.3~T. 
This possibility was also suggested by the fact 
that the scaled data of $\phi_\sigma$ at $B$=1~T were more scattered 
than those at $B$=0~T, leading to poorer fitting results of $\omega_0$ 
and $\sigma_0$ to the 3D-$XY$ model. 
As one possible candidate, we suspect that the vortex dynamics dominates the data.
Indeed, the estimated $\xi(T)$$\sim$100~nm for the 2D-``U" critical charge dynamics at 1.02$T_c$ 
exceeds $\ell_v$$\sim$80~nm at 0.3~T.
However, we found that the scaled data of $\phi_\sigma$ for $x$=0.17 above 0.3~T 
was qualitatively different from those observed for $x$=0.07 at 1~T, 
as shown in Figs.~\ref{fig:scaling07H} and \ref{fig:scaling17H}. 
In addition, if we use the 3D-XY critical exponent, 
$\xi(T)$ in the measurement temperature range is estimated at most $\sim$60~nm, 
which is smaller than $\ell_v$$\sim$80~nm at 0.3~T.
Therefore, there is another possibility that the critical charge dynamics 
changed from the 2D-``U" to the 3D-$XY$ universality class, 
even though it remains unclear at present why the critical charge dynamics for $x$=0.17 
was largely dependent on the applied magnetic field. 

In summary, 
the systematic studies of the effects of the external magnetic field on the critical charge dynamics 
confirmed that our determination of the universality class, 
which was classified into 2D-$XY$ for the underdoped region, 3D-$XY$ for the optimally doped region, 
and 2D-``U" for the overdoped region, was plausible. 

\section{Discussion}
As was already pointed out, 
we discovered three different universality classes in the phase diagram of LSCO, 
representing the existence of two kinds of crossover lines across the phase diagram. 
One is the dimensional crossover from 2D-$XY$ to 3D-$XY$ between $x$=0.14 and $x$=0.15, 
while the other is the crossover from 3D-$XY$ to 2D-``U" between $x$=0.16 and $x$=0.18. 
The study of the effects of thickness and magnetic field dependence supported 
that these features were not caused by the finite size effects, 
but were intrinsic to the superconducting phase transition in LSCO. 
In this section, 
we discuss possible origins of these changes of the universality class 
with the Sr concentration from various points of views.

\subsection{\label{sec:2D-XY}The 2D-$XY$ critical charge dynamics in the underdoped region} 
\begin{figure}
\includegraphics[width=0.9\linewidth]{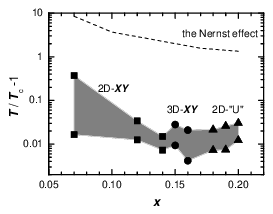}
\caption{
Plots of the highest (or the lowest) temperatures 
where a critical slowing down in $\omega_0(T)$ was observed as a function of the Sr concentration (hatched area).
Difference in symbols represents the difference in the observed critical behavior, 
which is classified into three groups of 2D-$XY$ (squares), 3D-$XY$ (circles), and 2D-``U"(triangles). 
Below the dashed line, the Nernst signal was observed.~\cite{Nernst}
} 
\label{fig:phase1} 
\end{figure}

\begin{figure}
\includegraphics[width=\linewidth]{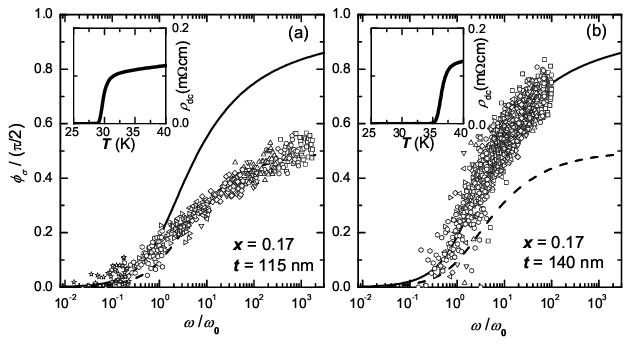}
\caption{
Scaled data of $\phi_{\sigma}$ for the $x$=0.17 films with different thicknesses.
They are plotted in a temperature range from $T$=29.0~K to 30.8~K for the $t$=115~nm film, 
and from $T$=35.0~K to 36.6~K for the $t$=140~nm film.
Solid (dashed) lines are the 2D (3D) Gaussian scaling function. 
Insets: Temperature dependence of dc resistivity of the corresponding films.
}
\label{fig:scaling17t}
\end{figure}

\begin{figure}
\includegraphics[width=\linewidth]{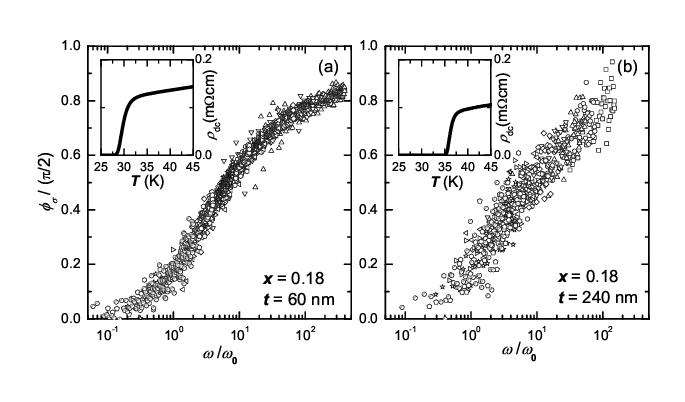}
\caption{
Scaled data of $\phi_{\sigma}$ for the $x$=0.18 films with (a) 60~nm thick and (b) 240~nm thick.
They are plotted in a temperature range from $T$=28.2~K to 31.0~K for the $t$=60~nm films, 
and from $T$= 35.0~K to 37.0~K for the $t$=240~nm films.
Insets: Temperature dependence of dc resistivity of the corresponding films.
}
\label{fig:scaling18t}
\end{figure}

\begin{figure}
\includegraphics[width=0.8\linewidth]{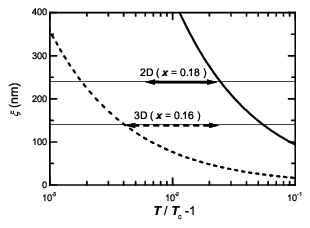}
\caption{
Dashed (solid) thick line represents a critical divergence of $\xi(T)=\xi_0(T/T_c-1)^{0.67}$ 
in the 3D-$XY$ critical charge dynamics with $\xi_0$ =3.5~nm (20~nm). 
Dashed and solid arrows show temperature ranges
where the critical behavior was observed for the two LSCO ($x$=0.16 and $x$=0.18, respectively). 
Thin solid lines show the thicknesses, $t$, of the two films. 
Note that the 3D-$XY$ critical charge dynamics of the $x$=0.16 film
was not cut off by the film thickness ($t$=140~nm) in the observed temperature range. 
This suggests that $\xi_0$ is smaller than 3.5~nm. 
On the other hand, if the 2D critical behavior of the $x$=0.18 film in the observed temperature range 
was attributed to the 3D-$XY$ critical behavior, 
which was cut off by the film thickness ($t$=240~nm), 
$\xi_0$ must be larger than 20~nm. 
}
\label{fig:xic-thickness}
\end{figure}

\begin{figure*}
\includegraphics[width=0.8\linewidth]{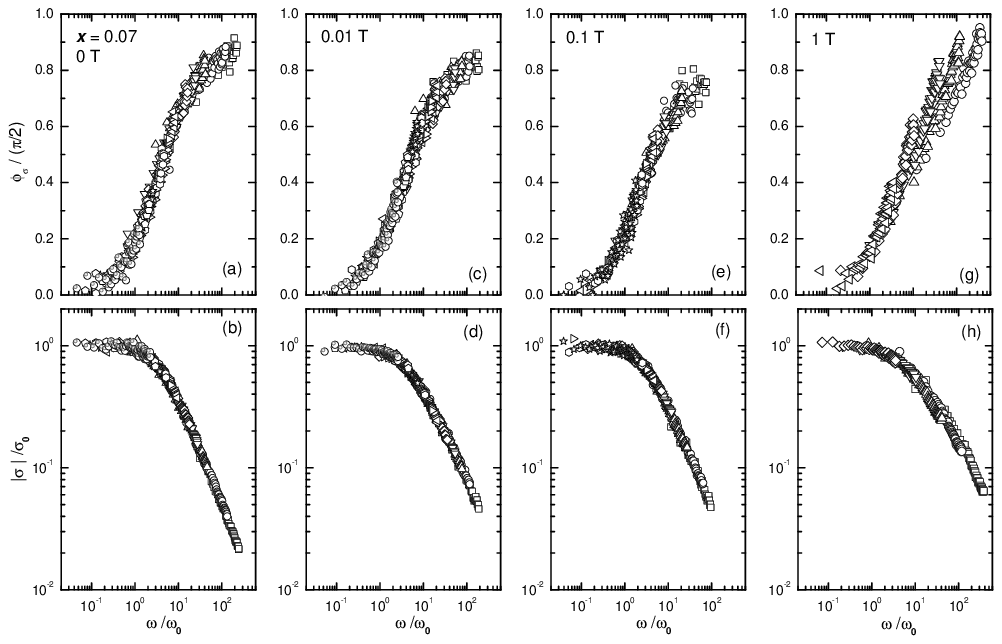}
\caption{
Scaled data of $\phi_{\sigma}$ (upper panels) and $|\sigma|$ (lower panes) 
for the $x=0.07$ film in zero and finite magnetic fields. 
}
\label{fig:scaling07H}
\end{figure*}

\begin{figure*}
\includegraphics[width=0.4\linewidth]{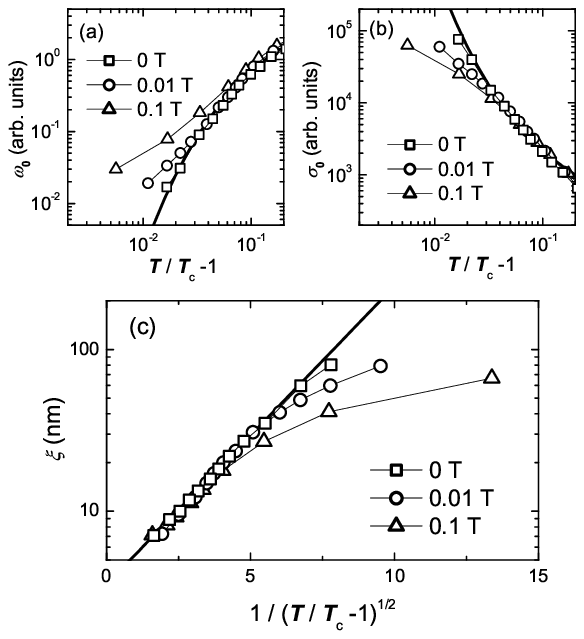}
\caption{
Temperature dependence of (a) $\omega_0$, (b) $\sigma_0$, and (c) $\xi\propto1/\sqrt{\omega_0}$ for the $x=0.07$ film. 
Solid lines in each panel are calculations given by Eq.~(\ref{xiKT}) with $\xi_0$=3~nm. 
}
\label{fig:para07H}
\end{figure*}

\begin{figure*}
\includegraphics[width=0.9\linewidth]{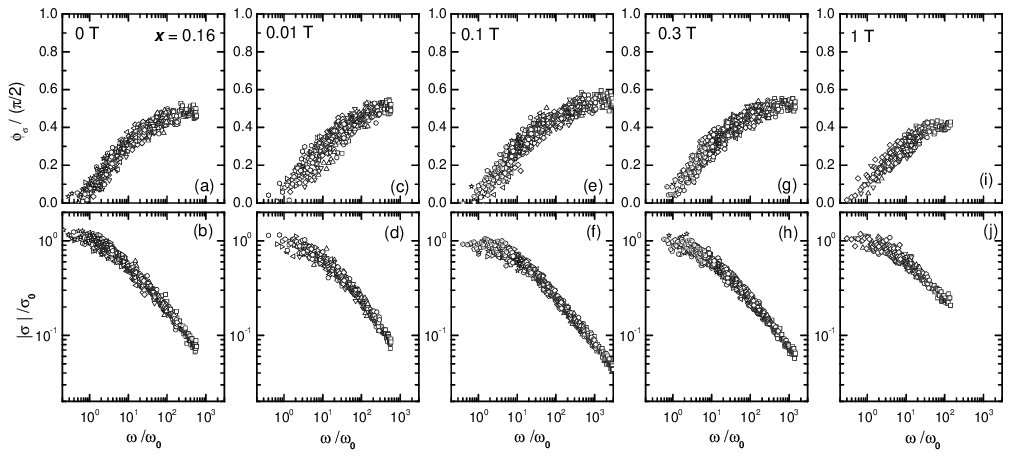}
\caption{
Scaled data of $\phi_{\sigma}$ (upper panels) and $|\sigma|$ (lower panes) 
for the $x$=0.16 film in zero and finite magnetic fields.
}
\label{fig:scaling16H}
\end{figure*}

\begin{figure*}
\includegraphics[width=0.3\linewidth]{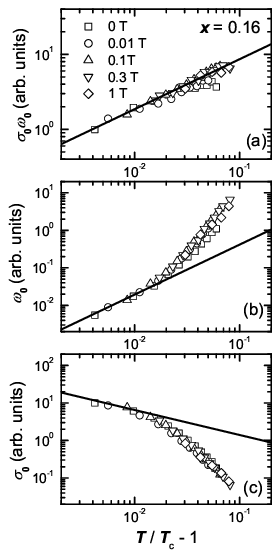}
\caption{
Temperature dependences of the obtained scaling parameters, 
(a) $\omega_0\sigma_0$, (b) $\omega_0$, and (c) $\sigma_0$, for the $x$=0.16 film 
in zero and finite magnetic fields. 
Solid lines are calculations given by Eqs.~(2) and (3) with $\nu$=0.67, $z$=2, and $d$=3, 
suggesting the 3D-$XY$ critical charge dynamics.
}
\label{fig:para16H}
\end{figure*}

\begin{figure*}
\includegraphics[width=0.9\linewidth]{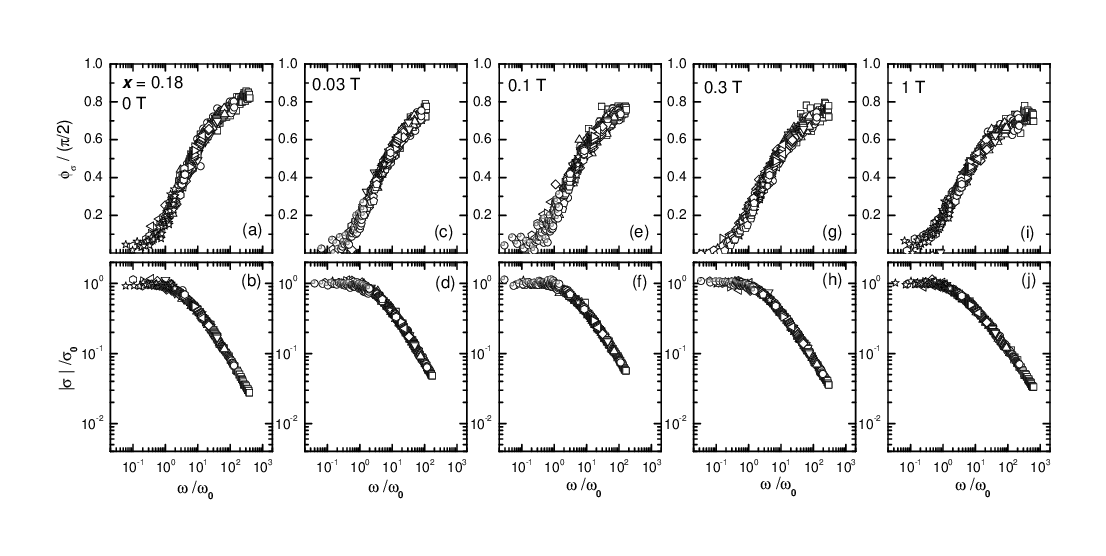}
\caption{
Scaled data of $\phi_{\sigma}$ (upper panels) and $|\sigma|$ (lower panes) 
for the $x$=0.18 film in zero and finite magnetic field.
}
\label{fig:scaling18H}
\end{figure*}

\begin{figure*}
\includegraphics[width=0.3\linewidth]{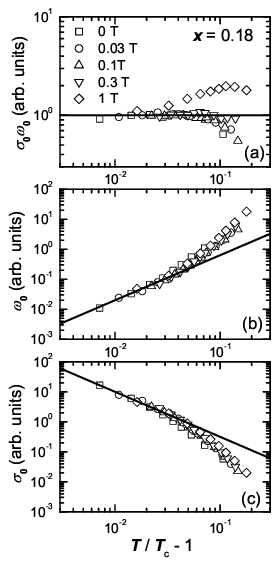}
\caption{
Temperature dependences of the obtained scaling parameters, 
(a) $\omega_0\sigma_0$, (b) $\omega_0$, and (c) $\sigma_0$, for the $x$=0.18 film 
in zero and finite magnetic fields. 
Solid lines are calculations given by Eqs.~(2) and (3) with $\nu z$=1.5 and $d$=2.
}
\label{fig:para18H}
\end{figure*}

\begin{figure}
\includegraphics[width=\linewidth]{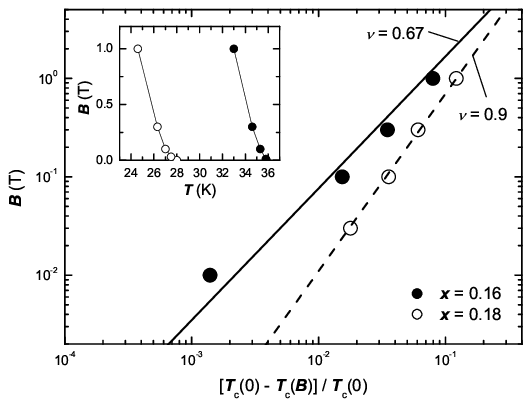}
\caption{
Plots of the applied magnetic field for the optimally doped LSCO ($x$=0.16) and the overdoped LSCO ($x$=0.18) 
as a function of $1-T_c(B)/T_c(0)$. 
Solid and dashed lines are calculations given by Eq.~(\ref{eq:Bc2}) 
with $\nu$=0.67 and $\nu$=0.9, respectively. 
Inset: Plots of the applied magnetic field, $B$, as a function of the critical temperature, $T_c(B)$ for each film, 
which was determined through the dynamic scaling analysis at each magnetic field. 
If vortex dynamics can be neglected, this plot agrees with the temperature dependence of $B_{c2}(T)$. 
} 
\label{fig:nu} 
\end{figure}

\begin{figure*}
\includegraphics[width=\linewidth]{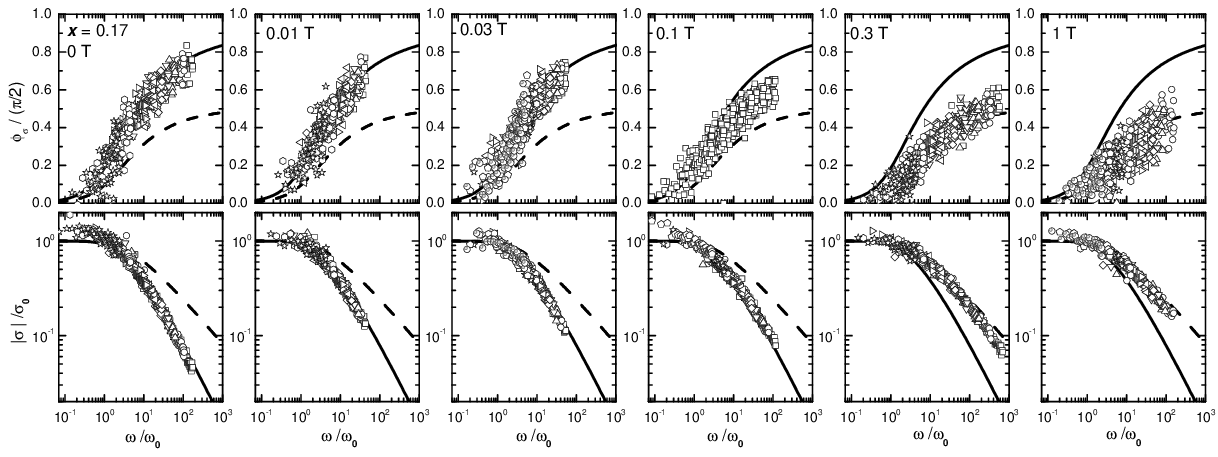}
\caption{
Scaled data of $\phi_{\sigma}$ (upper panels) and $|\sigma|$ (lower panels) 
for the $x$=0.17 film in zero and finite magnetic fields.
Solid (dashed) lines are the 2D (3D) Gaussian scaling functions.
}
\label{fig:scaling17H}
\end{figure*}

\begin{figure*}
\includegraphics[width=0.3\linewidth]{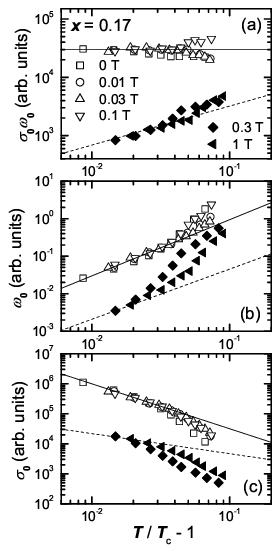}
\caption{
Temperature dependences of the obtained scaling parameters, 
(a) $\omega_0\sigma_0$, (b) $\omega_0$, and (c) $\sigma_0$, for the $x$=0.17 film 
in zero and finite magnetic fields. 
Solid (dashed) lines are calculations given by Eqs.~(2) and (3) with $\nu z$=1.5 and $d$=2 ($\nu$=0.67, $z$=2, and $d$=3).
}
\label{fig:para17H}
\end{figure*}

First, we discuss the 2D-$XY$ critical charge dynamics in the underdoped region ($x=0.07\sim 0.14$). 
This possibility was already pointed out by our previous study.~\cite{KitanoPRB2006} 
In this work, we confirmed that the exponential divergence of $\xi_{\rm KT}(T)$ for $x$=0.07 
was suppressed by applying small magnetic field, in agreement with the BKT theory 
which considered the effects of externally induced vortices.~\cite{Doniach} 
These new results strongly support that the critical charge dynamics in the underdoped region 
was classified into the 2D-$XY$ universality class. 
What is most significant is that the phase fluctuations of the superconductivity order parameter, 
which could be directly determined in our methods,~\cite{KitanoM2S} 
survived only up to much lower temperature (at most $\sim$1.4$T_c$) 
than the closing temperature of the pseudogap. 
As shown in Fig~\ref{fig:phase1},
this shows a sharp contrast to the Nernst experiments by Ong and Wang.~\cite{Nernst}
Our results indicated that the anomalous pseudogap phenomena cannot be explained 
only by the classical phase fluctuations of superconducting orders, 
in contrast to the early prediction by Emery and Kivelson.~\cite{EK95} 
A similar conclusion was also obtained by a theoretical consideration on ``cheap" vortices 
where the vortex core energy is very small ($\sim k_{\rm B}T_c$) and both of the amplitude 
and phase fluctuations are controlled by the same energy scale, $k_{\rm B}T_c$.~\cite{PALee2006} 

\begin{figure}
\includegraphics[width=0.8\linewidth]{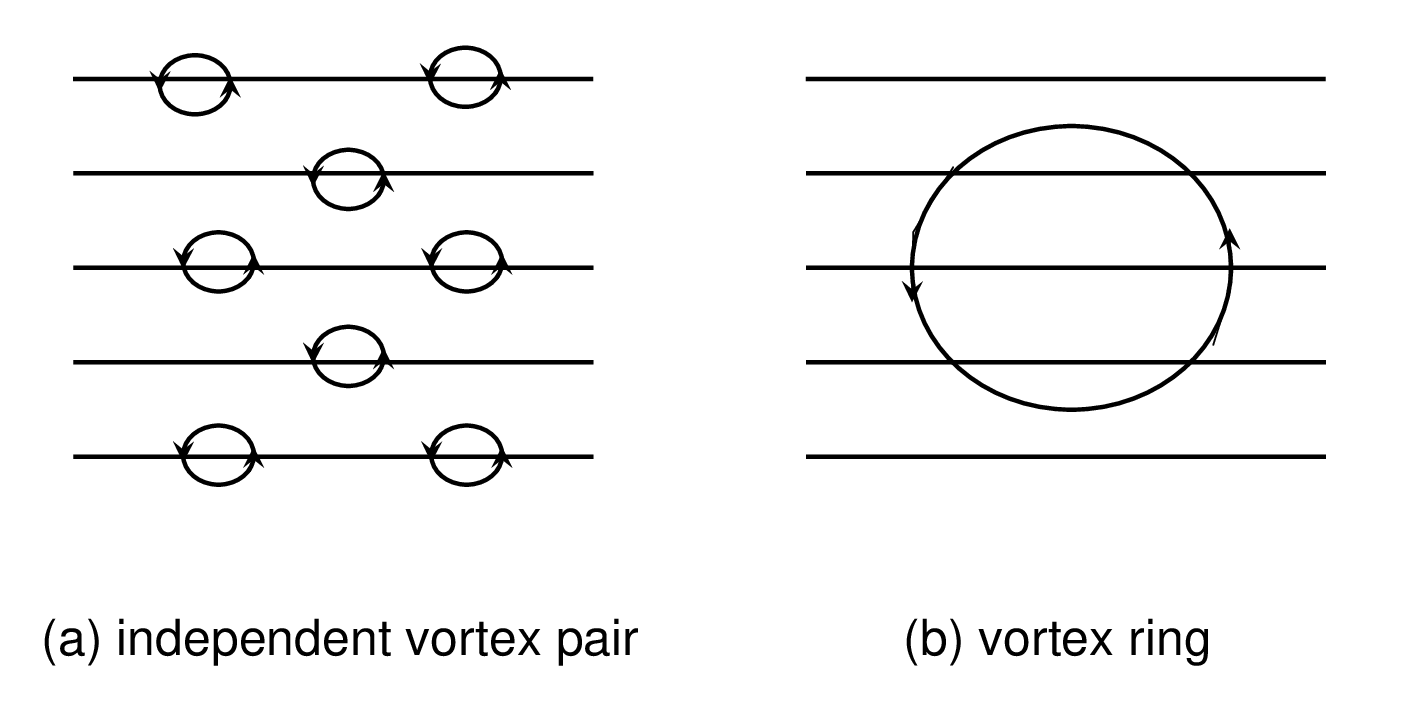}
\caption{
Schematic drawings of (a) independent vortex-antivortex pair (b) vortex ring in a layered superconductor. 
}
\label{fig:v_ring}
\end{figure}

In the layered superconductor, it is suggested that the finite interlayer coupling effect 
enhances the true critical temperature, $T_c$, as follows. 
\begin{equation}
T_c/T_{\rm BKT}=1+\Big(\frac{\pi}{\ln\sqrt{1/\Delta}}\Big)^2, 
\label{eq:TcTbkt}
\end{equation}
where $T_{\rm BKT}$, $\Delta$ are the BKT transition temperature 
and the anisotropy ratio of the interlayer coupling to the intralayer coupling, 
respectively.~\cite{HikamiTsuneto,Matsuda93} 
According to Eq.~(\ref{eq:TcTbkt}), 
the 2D-$XY$ critical behavior is no longer observed in the vicinity of $T_c$ 
and that the system should exhibit the 3D critical behavior, 
except for the case that the interlayer coupling is negligibly small ($\Delta\sim$~0). 
However, it should be pointed out that Eq.~(\ref{eq:TcTbkt}) overestimated $T_c$. 
In fact, Eq.~(\ref{eq:TcTbkt}) was derived by considering the critical radius, $r_c$, 
where the energy of the independent vortex pair balances with that of the vortex ring 
(See Fig.~\ref{fig:v_ring}). 
In Refs.~\onlinecite{HikamiTsuneto} and~\onlinecite{Matsuda93},
$r_c$ was estimated by neglecting the contribution of the vortex core energy, $E_c$, 
compared with the logarithmic interaction term. However, this is incorrect, 
which was already pointed out in the original paper of the BKT theory,~\cite{KT} 
and also in a recent theoretical review paper more clearly.~\cite{PALee2006} 
Particularly, according to the latter paper, 
$E_c$ is expected to be at least comparable to the logarithmic interaction term 
even for ``cheap" vortices. 
Thus, it is clear that Eq.~(\ref{eq:TcTbkt}) overestimates 
the enhancement of $T_c$ owing to the interlayer coupling effect. 
In order to discuss the possibility of the vortex ring excitation 
in the layered cuprate superconductor more precisely, 
we need the accurate estimation of $E_c$ of ``cheap" vortices, which remains an open issue. 

Even if we consider a somewhat unrealistic situation where $E_c$ is negligible, 
our results merely suggest that the interlayer Josephson coupling is not yet developed 
at least above the minimum temperature where the critical behavior 
was successfully observed ($\sim 1.01T_c$ for $x$=0.07, as shown in Fig.~\ref{fig:phase1}). 
Indeed, we found that the effective thickness of the superconducting sheet was comparable 
to the distance between the CuO$_2$ planes, 
suggesting that the fluctuating superconductivity appeared in each CuO$_2$ plane 
which remained nearly decoupled within the measured temperature range.~\cite{KitanoPRB2006} 
We emphasize that our results do not rule out the development of the interlayer Josephson coupling, 
which will finally lead to the 3D critical behavior. 
The Josephson coupling is expected to be developed in the further vicinity of $T_c$, 
which we could not approach experimentally. 
Thus, our observation of the 2D critical behavior 
does not contradicts the general property of a layered superconductor. 

\subsection{\label{sec:crossover}Two crossovers in the phase diagram} 
The most important findings are the two crossovers, that is, 
the crossover from 2D-$XY$ to 3D-$XY$ near $x$=0.14~\cite{KitanoPRB2006} 
and the other crossover from 3D-$XY$ to 2D-``U" near $x$=0.16 in the phase diagram.
First, we focus on the first crossover.
The sudden change of the universality class from the 2D-$XY$ to 3D-$XY$ 
around at $x$=0.14 is discussed in terms of several theoretical models.

In the $U$(1) mean field theory of the $t-J$ model,~\cite{Suzumura1988,PALee1992} 
$T_c$ is controlled by the Bose condensation of holon in the underdoped region, 
and the fermion pairing in the overdoped region. 
This means that 
a qualitative change of the condensation nature is expected to take place at the optimal doping.
Our results were consistent with this picture in the sense that 
the critical charge dynamics changes near the optimally doped region.
However, it seems to be difficult to explain the change of the dimensionality.
In the theory, 
the weak 3D interlayer hopping is assumed 
to obtain a finite transition temperature of the Bose condensation of holons. 
As was suggested by Lee and co-workers,~\cite{PALee2006} 
the BKT transition for the fermion pairing and the boson condensation can occur simultaneously 
where the gauge field becomes massive due to the Higgs mechanism. 
Thus, the phase transition lines for the fermion pairing and the boson condensation, 
which were predicted in the $U$(1) mean filed theory of the $t-J$ model, become the crossover lines 
and only the superconducting transition remains to be the real BKT transition. 
This suggests that we should observe only the 2D-$XY$ universality class 
in the whole range of the phase diagram. 
In other words, 
the anomalous change of the dimensionality in the critical behavior with carrier doping 
cannot be explained only by the early framework in the $t-J$ model 
suggesting the fermion pairing and the boson condensation. 
The so-called confinement-deconfinement transition 
due to the strong coupling gauge field~\cite{PALee2006,Nagaosa2000} 
is another candidate for the change of the critical behavior. 
Although a confinement-deconfinement transition does not always imply the change of the dimensionality, 
it is possible that a 3D Fermi liquid phase is realized 
as a consequence of the confinement of fermions and bosons, 
in which the 3D-XY critical behavior arises.

On the other hand, in the hidden QCP scenario, 
this sudden change of the universality class can be understood as a classical-quantum crossover 
which may occur with $x$ approaching the hidden QCP located beneath the superconducting dome.
A spatial and time correlation of the quantum fluctuation of the hidden order is strong 
near the QCP.
Therefore, provided that the quantum fluctuation couples to the thermal fluctuation of the superconducting order,
the critical dynamics around the superconduting transition 
can be affected by the quantum-fluctuation effects in the vicinity of the QCP.
In this picture, 
it is also predicted that the quantum fluctuation effects should become less prominent 
as $x$ increases away from the hidden QCP.
From this point of view, the behavior in the overdoped region reported in this paper 
is very crucial to examine these theoretical models.

In the overdoped region,
we found the second crossover from 3D-$XY$ to 2D-``U" near $x$=0.16, 
in addition to the first crossover near $x$=0.14.
This is highly unexpected, 
since the general trend is that the transport properties of the high-$T_c$ cuprates become less
anisotropic with carrier doping.~\cite{CooperReview}
The $t-J$ model based interpretation encounters a difficulty, 
since it does not expect any further dimensional crossover to take place. 
Even if a 3D Fermi liquid phase was realized by a confinement of fermions and bosons,
there is no reason why the 3D phase is realized 
only in a narrow range near the optimally doped region.
On the other hand, 
the theoretical models assuming the hidden QCP can naturally explain our finding
that the critical charge dynamics at least down to $T$$\sim$1.01$T_c$ was essentially 2D 
all over the phase diagram except for a narrow range near the optimally doped region.
This result was in correspondence with the expectation 
that the quantum fluctuation effects are prominent only in the vicinity of the QCP.

\begin{figure}
\includegraphics[width=\linewidth]{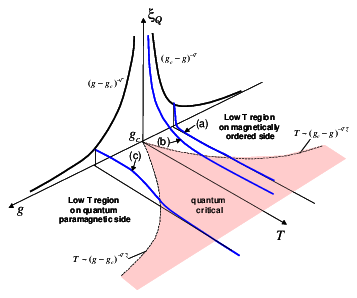}
\caption{
Schematic phase diagram of the 2D quantum antiferromagnet. 
Lines (a), (b) and (c) represent $\xi_Q$ in each region.
}
\label{fig:QPT}
\end{figure}

More precisely, the quantum fluctuation effects at finite temperatures are 
prominent only in the {\it quantum critical region}, 
which will be described below.
In general, in the 2D-QPT scenario, 
the hidden order of the QPT becomes long-range only at zero temperature.~\cite{SachdevTextbook} 
For instance, suppose the Heisenberg antiferromagnet with the 2D square lattice, 
which corresponds to an undoped CuO$_2$ plane.~\cite{Chakravarty1989} 
The magnetic long-range order is present only for $g<g_c$ at $T=0$, 
where $g$ and $g_c$ are a key parameter to control a QPT 
and a critical parameter corresponding to the QCP, respectively. 
At $T=0$, the correlation length $\xi_Q$ of QPT diverges as $\sim |g-g_c|^{-\tilde{\nu}}$ 
in the vicinity of $g_c$, where $\tilde{\nu}$ is a static quantum critical exponent. 
On the other hand, at finite temperatures, there are three characteristic regions, 
each of which is separated by two crossover lines, $T=\pm(g-g_c)^{\tilde{\nu}\tilde{z}}$, 
where $\tilde{z}$ is a dynamic quantum critical exponent, as shown in Fig.~\ref{fig:QPT}. 
Here, we used the same description for each region as that proposed by Sachdev,~\cite{SachdevTextbook} 
since the commonly used description such as ``quantum disordered" seemed to be somewhat misleading. 

In the {\it low $T$ region on the magnetically ordered side} [$0<T<(g_c-g)^{\tilde{\nu}\tilde{z}}$], 
both of $\xi_Q$ and the correlation time, $\tau_Q$, decrease exponentially as $T$ increases, 
suggesting that the quantum fluctuation effect 
is extremely small (line(a) in Fig.~\ref{fig:QPT}).
Thus, in this region, the system is regarded as classical at finite temperatures. 
In the {\it quantum critical region} 
[$T>(g_c-g)^{\tilde{\nu}\tilde{z}}$ and $T>(g-g_c)^{\tilde{\nu}\tilde{z}}$], 
$\xi_Q$ and $\tau_Q$ decrease as $\sim 1/T$ (line(b) in Fig.~\ref{fig:QPT}). 
This means that large quantum fluctuations survives even at high temperature 
which might appear in various properties, as well as thermal fluctuations.
Finally, in the {\it low $T$ region on the quantum paramagnetic side} 
[$0<T<(g-g_c)^{\tilde{\nu}\tilde{z}}$], 
$\xi_Q$ exponentially reaches a constant value as $T\to 0$ (line(c) in Fig.~\ref{fig:QPT}). 
The value of $\xi_Q$ at $T=0$ behaves as $|g-g_c|^{\tilde{\nu}\tilde{z}}$.
This region can also be regarded as a ``spin liquid" state,~\cite{PALee2006} 
since there is no long-range order even at $T=0$. 
In this region, although quantum fluctuations surviving at $T=0$ 
contribute to the oscillatory spin correlation, 
the long-range time correlation is exponentially relaxed by thermal fluctuations. 
Thus, the behavior of $\tau_Q$ at $T>0$ is rather dominated by the classical relaxation. 
In summary, the schematic behavior of $\xi_Q$ for the 2D Heisenberg antiferromagnet 
is shown in Fig.~\ref{fig:QPT}. 

\begin{figure}
\includegraphics[width=0.8\linewidth]{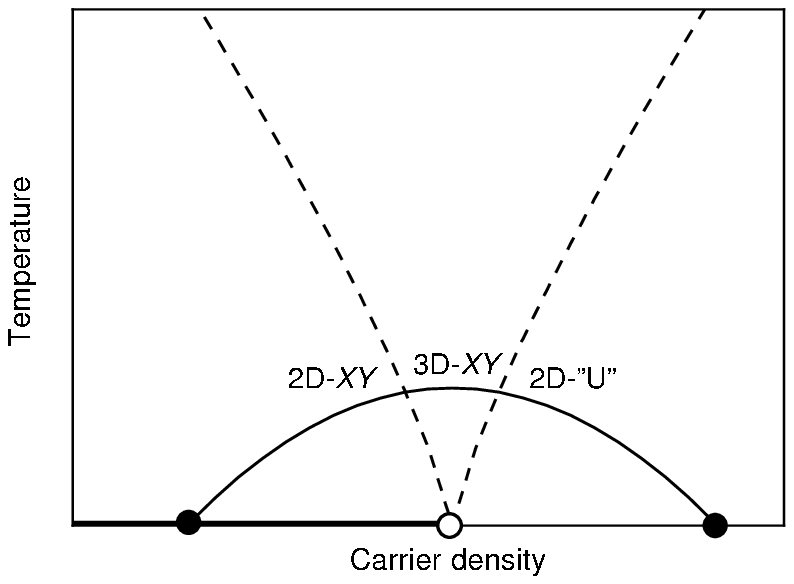}
\caption{
Schematic phase diagram of LSCO together with our results about the critical charge dynamics. 
Solid circles are the phenomenologically well-established QCPs. 
An open circle near the optimal doping is a hidden QCP 
assumed to explain our findings. 
The region between two dashed lines is the {\it quantum critical region}. 
Thick solid line at $T$=0 is a long range ordered region. 
}
\label{fig:phase2}
\end{figure}

When this concept is applied to discuss the phase diagram of the cuprates,
the parameter, $g$, is related to the carrier concentration, $n$.
As holes are doped into the CuO$_2$ plane at $T=0$, 
it is expected that the long-range order of the spin correlation 
is broken at a critical value of the hole doping, $n_c$, which is corresponding to $g_c$. 
Thus, one can consider that $g$ is a positive function of the hole doping and that $g(n_c)=g_c$. 
In the quantum critical region around $n_c$, 
the antiferromagnetic quantum fluctuation is considered to decay slowly as $\sim 1/T$.
It is already well-established that the antiferromagnetic quantum fluctuation 
favors the $d$-wave superconductivity.~\cite{Moriya,Pines} 
This suggests the possibility that a spatial and time correlation of the superconducting order 
can be enhanced by the quantum fluctuation effects.
Provided that the interlayer correlation is also enhanced, 
the observed dimensional crossover from 2D to 3D can be understood 
as a consequence of this quantum-fluctuation effects.
Note that, this effect is limited to the quantum critical region. 
In other two regions, 
$\tau_Q$ for the quantum fluctuation decays exponentially with increasing temperature, 
suggesting a purely classical dynamics. 
Thus, the experimental result that the 3D behavior was observed only in the vicinity of the optimally doped region
showed good qualitative agreement with this QCP scenario, as shown in Fig. \ref{fig:phase2}.

Note that the 2D antiferromagnetic order discussed here 
is just one of the candidates for the possible QPT. 
The scenario is essentially based on a hidden order in the 2D system.
The 2D order is present only at $T=0$, 
and becomes short-range at finite temperatures, independent of $g$.
Therefore, no experiment at finite temperatures can determine $n_c$ directly. 
This means that the disappearance of the 3D N\'{e}el order by hole-doping 
does not always imply the position of the QCP in the 2D antiferromagnet. 
Rather, a scaling analysis of the wave vector- and frequency-dependent magnetic fluctuations 
seems to be useful to search the position of the QCP. 
Indeed, the neutron scattering study using an LSCO single crystal suggested 
that the QCP might be hidden near $x$=0.14,~\cite{Aeppli1997} which agrees well with our results.

At present, this scenario seems to be compatible with our results most successfully.
The important requirement is 
that the quantum fluctuation around the hidden QCP couples to the thermal fluctuations 
of the superconducting order near the superconducting transition. 
Thus, a crucial question is what is the most plausible candidate for a hidden order 
among various proposals.~\cite{Moriya,Pines,Castellani1995,Varma1997,Chakravarty2001,Zaanen,KivelsonRMP2003,Vojta2000} 
A very recent discovery,~\cite{Kohsaka2007} of the 4$a_0$ wide unidirectional electronic domain 
which was detected by using a tunneling-asymmetry imaging technique in STM, 
may be also a candidate for such hidden orders. 
Unfortunately, we could not find a clear answer to this question in this work. 
In addition, there is a more complicated problem in the neighborhood of the Mott insulator-to-metal transition.
It is predicted that the coupling between the order parameter of the QPT 
and the low-energy fermions provides a more complex behavior 
with two dynamic quantum critical exponents.~\cite{Vojta2003,Rosch2001,Imada2005} 
In LSCO single crystals, the insulator-to-metal crossover was suggested to occur near $x$=0.16, 
by the in-plane resistivity measurements 
down to $T$=0.65~K using a 61~T pulse magnetic field.~\cite{Boebinger1996} 
An open issue is whether this behavior is attributed to the Mott transition or not. 

\subsection{\label{sec:2D-U}Possible origin on the 2D-``U" critical charge dynamics} 
The critical charge dynamics observed in the overdoped region 
was found to be different from the well-known universality classes 
(2D-$XY$, 3D-$XY$, or the Gaussian fluctuation). 
Here, we discuss this 2D-``U" critical behavior in details. 

\begin{figure}
\includegraphics[width=0.8\linewidth]{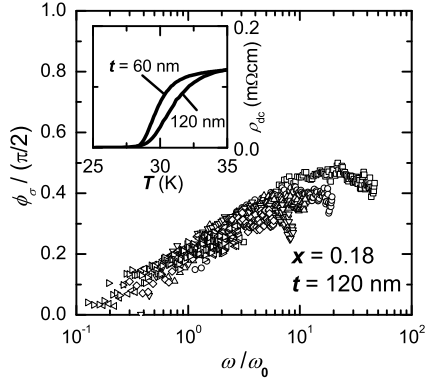}
\caption{
An example of the breakdown of the dynamic scaling 
for the $\phi_{\sigma}$ of the thicker film ($t$=120~nm) with $x$=0.18.
Inset: Temperature dependence of dc resistivity of the film 
compared with that of the thinner film ($t$=60~nm).
}
\label{fig:scalefail}
\end{figure}

First of all, we consider the effect of disorder due to the Sr substitution, 
which is expected to be more prominent with increasing Sr concentration. 
As was emphasized repeatedly, the successfully scaled data confirmed 
that the observed 2D-``U" critical behavior could not be explained by the distribution of $T_c$ 
due to the disorder effect, 
because our method to obtain the data collapse of $\phi_\sigma$ and $|\sigma|$ 
is sensitive to the breakdown of the dynamic scaling hypothesis, 
which can be caused by a broad distribution of $T_c$ due to disorders. 
Figure~\ref{fig:scalefail} shows an example of the breakdown of the dynamic scaling 
due to the distribution of $T_c$.
This is $\phi_\sigma(\omega)$ scaled by $\omega_0$ of the thicker film ($t$=120~nm) with $x=0.18$.
As shown in the inset of Fig.~\ref{fig:scalefail}, 
dc resistivity showed a broader transition width than that of the thinner film ($t$=60~nm), 
which yielded the successfully scaled $\sigma_{\rm fl}(\omega)$. 
The data of the thicker film does not show the dynamic scaling behavior, 
suggesting a broad distribution of $T_c$ due to disorders 
results in the breakdown of the dynamic scaling hypothesis.

In addition to such an experimental justification, 
the so-called Harris's criterion on the critical phenomenon also tells us 
that the weak disorder is irrelevant to the 2D-$XY$ or 3D-$XY$ critical behavior.~\cite{ref:Harris}. 
On the other hand, 
the strong disorder can change the critical behavior to a different one.
However, even for the strong disorder, 
it seems difficult to change the dimensionality in the critical charge dynamics. 
In other words, the observed dimensional crossover from 3D to 2D is a very robust one, 
which cannot be explained by the disorder effect.

Meanwhile, the strong disorder effect is one of the possible candidates 
for the origin of the anomalous critical exponents. 
The disorder effect on the 2D system has been investigated by the 2D random gauge $XY$ model, 
which introduced the disorder effect as the random variables in magnetic bond angles 
uniformly distributed in the range from $-r\pi$ to $r\pi$.~\cite{ref:Holme} 
When the disorder strength, $r$, is equal to 1, 
the model corresponds to the standard $XY$ gauge glass model,~\cite{ref:Huse} 
while the standard $XY$ model is recovered for $r=0$. 
Interestingly, a recent Monte Carlo simulation of the 2D random gauge $XY$ model suggested 
that a non-BKT type transition with $\nu\approx$ 1.1 occurred 
at low temperatures for a stronger disorder strength than a critical value of $r$, 
while the standard BKT transition occurred at low temperatures 
for a weaker disorder strength.~\cite{ref:Holme} 
The calculated critical exponent is not so different from our result ($\nu\sim$0.9) for $x$=0.18.
Thus, this is one possibility for the 2D-``U" universality class.
However, a better agreement of the critical exponent is obtained by considering 
the frustration effect on the 2D system.
We will discuss this later in terms of the frustrated 2D-$XY$ model.

Next, in the $t-J$ model assuming the BKT transition for the fermion pairing and the boson condensation, 
it is expected that the 2D-$XY$ critical behavior is observed in a whole region of the phase diagram. 
In practice, the BKT transition cannot be observed 
unless the BKT criterion, $\lambda_\perp\gg L$, is satisfied,~\cite{KT} 
where $\lambda_\perp(=\lambda^2/D_s)$ and $L$ is a screening length 
and a lateral size of a superconducting sheet, respectively, 
and $\lambda$ and $D_s$ is a bulk penetration depth 
and a thickness of the superconducting sheet, respectively. 
In the underdoped region, this criterion is satisfied, indeed.~\cite{KitanoPRB2006}
For overdoped films, 
$\lambda_\perp$ was estimated to be $10 \sim 100$~mm, which is sufficiently larger than $L\sim$ 1~mm, 
showing the criterion is satisfied also for overdoped samples.
Therefore, the 2D-``U" critical behavior was not induced by the failure of the BKT criterion.  
Thus, the 2D-``U" feature cannot be explained by the $t-J$ model based theories.

In a model assuming that a Fermi liquid state is realized in the heavily overdoped region, 
one may expect that a conventional Gaussian fluctuation rather than the critical fluctuation 
becomes more prominent with increasing the Sr concentration. 
From this point of view, 
a crossover from the 2D-$XY$ critical behavior to the Gaussian fluctuation can be a candidate 
for the origin of the observed anomalous critical exponents. 
In order to examine this, 
we estimated $1/\tau$ for each film ($\tau$ is the correlation time of the fluctuation), 
by comparing the scaled data of $\phi_\sigma(\omega)$ 
with the Gaussian scaling function.~\cite{Schmidt}
In the GL theory assuming the Gaussian fluctuation, 
it is well known that the divergence of $\tau$ near $T_c$ is given as follows. 
\begin{equation}
\tau_{\rm GL }=\frac{\pi \hbar}{16k_B|T-T_c|}.
\label{tauGL}
\end{equation}

\begin{figure}
\includegraphics[width=0.9\linewidth]{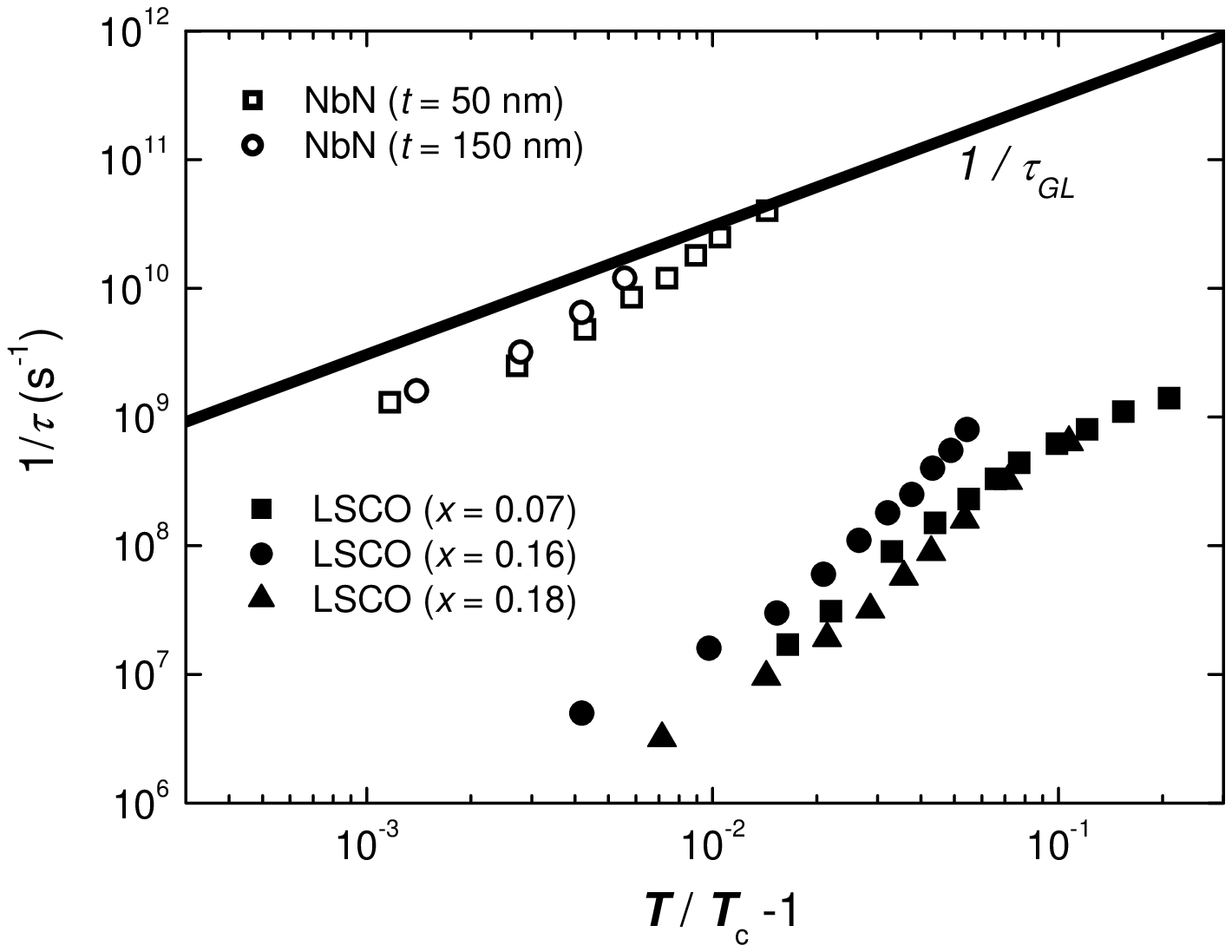}
\caption{
Temperature dependence of $1/\tau$, where $\tau$ is the correlation time. 
Solid symbols are the data of LSCO films ($x$=0.07, 0.16, 0.18) in this work, 
and open symbols are the data of NbN films ($t$=50~nm, 150~nm) in Ref.~\onlinecite{OhashiPRB2006}. 
A solid straight line is $1/\tau_{\rm GL}$ given by Eq.~(\ref{tauGL}). 
}
\label{fig:tauinv}
\end{figure}

Surprisingly, we found that all the results of $1/\tau$ for $x$=0.07, 0.16 and 0.18 
were much smaller than $1/\tau_{\rm GL}$ by two or three orders of magnitude, 
as shown in Fig.~\ref{fig:tauinv}. 
A similar slow critical charge dynamics has been reported 
for the nearly optimally doped YBa$_2$Cu$_3$O$_y$ thin film.~\cite{Booth96} 
These results show a sharp contrast to the similar plots of $1/\tau$ for NbN thin films,~\cite{OhashiPRB2006} 
which almost agreed with $1/\tau_{\rm GL}$ ( also shown in Fig.~\ref{fig:tauinv} ). 
Therefore, the observed dynamics is indeed attributed to 
the {\it critically slowing down} phenomenon in the critical region, 
and it cannot be explained by a crossover to the Gaussian fluctuation.
Interestingly, such a slow dynamics is also expected in the low $T$ regions 
both on the magnetically ordered and the quantum paramagnetic sides around the QCP, 
because of the opening of an energy gap in the excitation spectrum.~\cite{SachdevTextbook} 

In terms of the QCP scenario, 
we believe that the spin liquid state in the low $T$ region on the quantum paramagnetic side 
seems to provide key informations to understand the origin of the 2D-``U" critical behavior.
When a hole is doped in the CuO$_2$ plane, 
the spin on a Cu site is combined to the spin of the doped hole 
to form the Zhang-Rice (ZR) singlet,~\cite{ZR1988} 
giving rise to the spatial modulation of the nearest-neighbor exchange interaction, 
$J$, to a square lattice in the CuO$_2$ plane. 
In other words, the formation of the ZR singlet by doped holes 
introduces the frustration between the spins in the 2D antiferromagnet, 
as pointed out by Anderson.~\cite{Anderson1987} 
In particular, mobile ZR singlets can give rise to the dynamic spin frustration. 
This dynamic or static frustration effect has been considered to play an essential role 
in stabilizing the spin liquid or glass state.~\cite{Anderson1987,Edwards1975} 
Such a frustrated state might be related to the 2D-``U" critical behavior in the overdoped region.
Thus, we need to consider more explicitly the influence of the frustration effect 
to the superconductivity realized in the quantum paramagnetic side. 

The static frustration effect in the $XY$ spin systems has been extensively investigated 
by the so-called frustrated $XY$ model~\cite{ref:FFXY} 
both theoretically~\cite{ref:FFXYth} and experimentally.~\cite{ref:FFXYex} 
Interestingly, the study of the critical behavior in frustrated $XY$ spins 
seems to suggest the possibility of a new universality class, 
which is clearly different from the 2D-$XY$ universality class.~\cite{ref:FFXYth} 
Indeed, many numerical simulations for the fully frustrated $XY$ model 
suggested that $\nu$=0.8 - 1.0,~\cite{ref:FFXYth}, 
which shows a good agreement with our result ($\nu$$\sim$0.9) for $x$=0.18. 
Although the explicit correspondence between the frustrated $XY$ spin system 
and the superconductivity realized in the spin liquid state is an open issue to be resolved, 
the resemblance between the critical behaviors of both systems 
suggests the importance of frustration effect in the overdoped region.

\section{conclusion}
We investigated the critical charge dynamics of the superconducting to the normal-state transition 
for LSCO thin films with a wide range of the Sr concentration ($x$=0.07 to 0.20), 
through the dynamic scaling analysis of the fluctuation-induced microwave conductivity. 
We discovered that the critical charge dynamics in the phase diagram of LSCO 
was classified into the following three different universality classes; 
(i) the 2D-$XY$ universality class for the underdoped region ($x$=0.07 to 0.14), 
(ii) the 3D-$XY$ universality class for the nearly optimally doped region ($x$=0.15 and 0.16), 
(iii) the 2D-``U" ($\nu\sim$0.9, $z\sim$2) universality class for the overdoped region ($x$=0.17 to 0.20). 
We confirmed that the anomalous 2D critical behavior for the overdoped region 
is not induced by the finite size effect but is intrinsic to the overdoped LSCO.

Our results indicated that there are two kinds of dimensional crossovers in the phase diagram of LSCO, 
suggesting that the dimensionality in the critical charge dynamics changes twice with hole doping. 
In other words, 
except for the 3D-$XY$ critical charge dynamics observed only near the optimally doped concentration, 
we found that the critical charge dynamics of LSCO was essentially two-dimensional. 
This suggests that the fluctuating high temperature superconductivity initially appears 
in each CuO$_2$ plane, preceding the development of the interlayer Josephson coupling. 
Our results for the underdoped LSCO also indicated 
that the 2D-$XY$ critical charge dynamics could be observed at most up to $\sim1.4~T_c$, 
which was much lower than a closing temperature of the pseudogap. 
Thus, some of the early naive pictures about the pseudogap 
and the critical fluctuations in high-$T_c$ cuprates need to be modified. 

We discussed the possible origins of the anomalous dimensional crossovers.
It is concluded that neither the weakened anisotropic property with hole doping 
nor the confinement-deconfinement transition discussed in the $t-J$ model 
can explain the second dimensional crossover from the 3D-$XY$ to the 2D-``U", 
which occurs between $x$=0.16 and 0.17. 
On the other hand, the so-called hidden ``QCP" scenario, 
where the quantum critical fluctuation near the hidden QCP couples to the superconducting fluctuation, 
was able to explain the existence of two kinds of dimensional crossovers qualitatively. 
This scenario also seems to provide a key infromation 
to understand the origin of the 2D-``U" critical behavior, 
whose critical exponents were found to be similar to those of the frustrated $XY$ model.
The importance of the frustration effect in the spin liquid state is implied

At present, 
the scenario based on a hidden order of QPT in the 2D system seems to be the most reasonable candidate 
for the anomalous change of the universality class with hole doping. 
However, the hidden order which provides the direct evidence for QCP has not been yet discovered. 
If the hidden order is formed in the 2D system, 
it will be almost impossible to be found at finite temperatures. 
Thus, the dynamic scaling analysis including the quantum critical fluctuations 
plays an important role to settle this issue.

\section*{ACKNOWLEDGEMENTS}
We thank H. Fukuyama, Y. Matsuda, K. Fukushima for
fruitful discussions and comments.
This work was partly supported by 
the Grant-in-Aid for Scientific Research 
(13750005, 15760003, 17038006, 17340102, and 19014005) 
from the Ministry of Education, Science, Sports and Culture of Japan. 
T. Ohashi thanks 
the Japan Society for the Promotion of Science for financial support.

\newpage 


\clearpage

\clearpage

\end{document}